\documentclass[journal]{IEEEtran}

\ifCLASSINFOpdf
\else
\fi

\usepackage{cite}
\usepackage{afterpage}
\usepackage{verbatim} 
\usepackage{multicol}
\usepackage{array}
\usepackage{psfrag}
\usepackage[normalem]{ulem}
\usepackage{amsthm}
\usepackage{amsmath, amssymb}
\usepackage{amsfonts}
\usepackage{mathdots}
\usepackage{cite}
\usepackage{multirow}
\usepackage{setspace}
\usepackage{graphicx}
\usepackage{graphics}
\usepackage{booktabs}
\usepackage{subfig}
\usepackage{url}
\usepackage{verbatim} 
\usepackage{multicol}
\usepackage{array}
\usepackage{colortbl}
\usepackage{psfrag}
\usepackage{xcolor}
\usepackage{enumerate}
\usepackage{amsthm}
\usepackage{amsmath, amssymb}
\usepackage{amsfonts}

\usepackage{algorithm}
\usepackage{algorithmic}
\usepackage{dblfloatfix}

\usepackage{tikz}
\usetikzlibrary{automata,shapes,arrows,matrix,shadows,positioning,decorations.pathreplacing,decorations.text,trees,calc,decorations.markings,patterns}

\definecolor{mycolor}{RGB}{135,206,250} 

\tikzset{
	myellipse/.style args={[#1][#2]}{
		ellipse,
		anchor=north west,
		minimum width = #1em,
		minimum height = #2em,
		text centered,
		font=\large,
		text width = #1em - 0.5em,
		fill=mycolor
	},
}

\tikzset{
	myarrow/.style args={[#1][#2][#3]}{
		decoration={markings,
			mark=at position 0.5 with {\node[midway,anchor=#1,rotate=#2,text centered] {\footnotesize #3};},
			mark=at position 1 with {\arrow{>}},
		},
		postaction={decorate},
		shorten >=0pt,
		thick
	},
	myarrow/.default={[north][0][]},
}

\newcommand{\defeq}{{\stackrel{\Delta}{=}}}

\def\deg{\mathop{\rm deg}\nolimits} 
\def\qdeg{\mathop{\rm deg_q}\nolimits} 
\DeclareMathOperator*{\argmax}{arg\,max}
\DeclareMathOperator*{\argmin}{arg\,min}

\DeclareMathOperator{\wt}{wt}
\DeclareMathOperator{\rank}{rank}

\DeclareMathOperator{\sgn}{sgn}
\DeclareMathOperator{\diag}{diag}

\newcommand{\field}[1]{\mathbb{#1}}
\newcommand{\transpose}{\textsf{T}}

\theoremstyle{definition}
\newtheorem{thm}{Theorem}
\newtheorem{defn}{Definition}
\newtheorem{lem}{Lemma}

\newtheorem{prop}{Property}

\newtheorem{rem}{Remark}

\theoremstyle{plain}
\newtheorem{ex}{Example}

\newcommand{\bc}{\mathbf{c}}

\newcommand{\be}{\mathbf{e}}
\newcommand{\bff}{\mathbf{f}}

\newcommand{\br}{\mathbf{r}}
\newcommand{\bs}{\mathbf{s}}

\newcommand{\bx}{\mathbf{x}}
\newcommand{\by}{\mathbf{y}}

\newcommand{\bzero}{\mathbf{0}}

\newcommand{\bA}{\mathbf{A}}

\newcommand{\bC}{\mathbf{C}}
\newcommand{\bD}{\mathbf{D}}

\newcommand{\bF}{\mathbf{F}}
\newcommand{\bG}{\mathbf{G}}

\newcommand{\bM}{\mathbf{M}}

\newcommand{\bO}{\mathbf{O}}

\newcommand{\bR}{\mathbf{R}}

\newcommand{\bT}{\mathbf{T}}

\newcommand{\bX}{\mathbf{X}}
\newcommand{\bY}{\mathbf{Y}}

\newcommand{\bGEX}{\bG^{\mathrm{EX}}}

\newcommand{\isolatedchannel}{\mathcal{CH}}

\begin{document}
%
\title{Convolutional Codes with Maximum Column Sum Rank for Network Streaming}



\author{Rafid~Mahmood,~\IEEEmembership{Student Member,~IEEE}, Ahmed~Badr,~\IEEEmembership{Member,~IEEE}, and~Ashish~Khisti~\IEEEmembership{Member,~IEEE}
\thanks{R.~Mahmood, A.~Badr, and A.~Khisti (\{rmahmood, abadr, akhisti\}@comm.utoronto.ca) are with the University of Toronto, Toronto, ON, Canada.}
\thanks{Part of this work was presented at the 2015 International Symposium on Information Theory (ISIT) in
Hong Kong \cite{ref:mahmood_isit15}.}}

\maketitle

\begin{abstract}
The column Hamming distance of a convolutional code determines the error correction capability when streaming over a class of packet erasure channels. We introduce a metric known as the column sum rank, that parallels column Hamming distance when streaming over a network with link failures. We prove rank analogues of several known column Hamming distance properties and introduce a new family of convolutional codes that maximize the column sum rank up to the code memory. Our construction involves finding a class of super-regular matrices that preserve this property after multiplication with non-singular block diagonal matrices in the ground field.
\end{abstract}


%
\IEEEpeerreviewmaketitle

\section{Introduction}

In streaming communication, source packets arrive sequentially at the transmitter and are only useful for playback by the receiver in the same order. Erased packets must be recovered within a given maximum delay or be considered permanently lost. Streaming codes recover packets within these decoding deadlines and have previously been studied for single-link communication \cite{ref:martinian_it04, ref:leong_isit13, ref:layered_it15, ref:tomas_it12}. The works referenced in \cite{ref:martinian_it04, ref:leong_isit13, ref:layered_it15} focused primarily on low-delay recovery against burst losses, which are the predominant erasure patterns in Internet streams \cite{ref:ellis_pv12}. Alternatively, \cite{ref:tomas_it12, ref:layered_it15} considered coding for channels with arbitrary erasure patterns, restricting only the number of erasures in a window. It was shown that the column Hamming distance determines the maximum tolerable number of erasures that can occur in any window of the stream for decoding to remain successful. If there are fewer erasures than the distance in every sliding window, each source symbol is recovered within a given delay. A family of memory $m$ convolutional codes, known as $m$-Maximum Distance Separable ($m$-MDS) codes, attain the maximum column Hamming distance up to the code memory. Furthermore, these are constructed from block Toeplitz super-regular matrices\cite{ref:smds_it06, ref:hutch_laa08, ref:tomas_it12, ref:almeida_laa13}. These codes can also be used as constituent codes in the construction of optimal burst error correction codes for streaming systems \cite{ref:layered_it15}.

Suppose that a transmitter sends packets to several users through a series of intermediate nodes. Using generation-based linear network codes, the problem of decoding is reduced to inverting the channel transfer matrix between the transmitted and received packets \cite{ref:ahlswede_it00, ref:chou_allerton03, ref:ho_it06}. If links in the network fail to transmit packets for a given time instance, the rank of the channel matrix decreases, making packet recovery infeasible. One solution is to use end-to-end schemes to precode channel packets before transmission. Rank metric codes such as Gabidulin codes\cite{ref:mrd_pit85, ref:roth_it91} are capable of protecting packets in rank-deficient channels.  The minimum rank distance of a code determines the maximum permissible rank loss in a channel matrix. This method can be considered for single-shot network coding \cite{ref:kotter_it08, ref:silva_it08} or for multishot extensions \cite{ref:nobrega_winc10, ref:antonia_14}.

In this work we study a streaming setup that extends the single-link model of~\cite{ref:martinian_it04, ref:leong_isit13, ref:layered_it15} to a network. 
We assume that the intermediate nodes implement linear network coding and produce a channel transfer matrix relating the transmitted and received packets. To combat link failures, the source stream is further precoded at the source using a {\emph{streaming code}}. 
We define a new metric, the column sum rank and introduce a new family of convolutional codes that attain the maximum value for this metric. These are rank metric analogues of $m$-MDS codes \cite{ref:smds_it06}. Just as the column Hamming distance determines the maximum allowable number of erasures in single link streaming, we show that the column sum rank determines the maximum rank deficiency of the channel. Interestingly, there has been little prior work on rank metric convolutional codes. To our knowledge, the only previously studied construction appears in \cite{ref:antonia_14}, where the authors consider the active column sum rank as the metric of importance. Consequently, their approach and results differ from the present work both in the code constructions and applications.

This paper is outlined as follows.
The network streaming problem is introduced in Section \ref{sec:network_streaming_problem}. 
We provide an overview of rank metric block codes and $m$-MDS codes in Section \ref{sec:prior_codes}. 
The column sum rank is defined in Section \ref{sec:msr_thms}, where we derive several properties and establish their relevance in network streaming. 
Codes that maximize the column sum rank are referred to as Maximum Sum Rank (MSR) codes. 
We introduce a class of super-regular matrices in Section~\ref{sec:preservation} that preserve super-regularity after multiplication with block diagonal matrices in the ground field and use these to construct an MSR code in  Section \ref{sec:code_construct}. 
We conclude this paper with code examples and a discussion on the necessary field size.

\section{Network Streaming Problem}
\label{sec:network_streaming_problem}

The streaming problem is defined in three steps: the encoder, network model, and decoder. 
Encoding is performed in a causal fashion, as the incoming packets are not known until the time at which they must be transmitted. A linear network code has  been applied to the network and each node receives and sends linear combinations of the symbols in the channel packet. Consequently the network is abstracted to a channel matrix, which is assumed to be known to the receiver \cite{ref:ho_it06}. The decoder observes linear combinations of the symbols in each transmitted packet and must recover the source within the imposed deadline.

\subsection{Encoder}
Let $q$ be a prime power and $M \geq 0$. At each time instance $t \geq 0$, a source packet $\bs_t \in \field{F}_{q^M}^k$ arrives at the transmitter node. A channel packet $\bx_t \in \field{F}_{q^M}^n$ is constructed via a causal function of the previous source packets $\gamma_t(\bs_0,\dots,\bs_t)$. 
We consider the class of linear time invariant encoders.
A rate $R = \frac{k}{n}$ encoder with memory $m$ generates the channel packet
\begin{align}
\label{eq:network_encoder}
\bx_t = \sum_{i=0}^{m} \bs_{t-i} \cdot\bG_i,
\end{align}
using a set of generator matrices $\bG_i \in \field{F}_{q^M}^{n\times k}$ for $0 \leq i \leq m$.

\subsection{Network Model}
The transmitter node sends one channel packet through the network at each time instance. Although there is a natural delay in the end-to-end transmission due to link delays, we assume that such delays are sufficiently small so that one time instance contains the encoding, transmission, and decoding of a single channel packet\footnote{For example in audio streaming,  coded speech packets are generated once every $20$~ms. When the propagation delays are much smaller than this value, they can be ignored.}. The transmission of a single channel packet over one time instance is referred to as a \textit{shot}. 
 In each shot, the destination node observes $\by_t = \bx_t \bA_t$, where $\bA_t \in \field{F}_q^{n\times n}$ is the channel matrix at time $t$,
and is known to the receiver. In practice, the coefficients for the linear transformations applied by each node can be encoded into header bits, which the receiver uses to reconstruct the channel matrix \cite{ref:ho_it06}.  

Each shot is independent of all others. Communication over a window $[t,t+W-1]$ of $W$ shots is described using $\by_{[t,t+W-1]} = \bx_{[t,t+W-1]} \bA_{[t,t+W-1]}$, where $\bA_{[t,t+W-1]} = \diag{(\bA_t,\dots,\bA_{t+W-1})}$ is a block diagonal channel matrix\cite{ref:khisti_isit10, ref:nobrega_winc10}. Let $\rho_t \defeq \rank{(\bA_t)}$ denote the rank of $\bA_t$, for all $t \geq 0$. The sum of the ranks of the individual channel matrices is equal to the rank of the channel matrix in the window, i.e., $\sum_{i=t}^{t+W-1} \rho_i = \rank{(\bA_{[t,t+W-1]})}$.
Suppose that at any time instance, a link in the network may fail to transmit its intended symbol.
Intermediate nodes that do not receive an intended symbol simply do not include that symbol in the linear combination they transmit.
If all links are functional in the shot at any time $t$, then $\rho_t = n$, but failing links may result in a rank-deficient channel matrix at that time. 
One failing link can eliminate at most one of the min-cut paths connecting the transmitter and receiver. It follows that $\rank{(\bA_t)}$ is reduced by at most $1$ for every failing link \cite{ref:jalali_ita11}.
We introduce a sliding window model to characterize rank deficiencies in the network. 

\begin{defn}
Consider a network where for all $t\geq 0$, the receiver observes $\by_t = \bx_t \bA_t$, with $\rho_t \defeq \rank{(\bA_t)}$. The \textbf{Rank-Deficient Sliding Window Network} $\isolatedchannel(S,W)$ has the property that in any sliding window of length $W$, the rank of the block diagonal channel decreases by no more than $S$, i.e., $\sum_{i=t}^{t+W-1} \rho_i \geq nW - S$ for each $t \geq 0$. 
\end{defn}

In analysis, we disregard the linearly dependent columns of the channel matrix and the associated received symbols. At each time instance, the receiver effectively observes $\by^*_t = \bx_t \bA^*_t$, where $\bA^*_t \in \field{F}_q^{n \times \rho_t}$ contains only the linearly independent columns of $\bA_t$ and is referred to as the reduced channel matrix.

\begin{rem}
A sliding window model has been used in prior works on delay-constrained coding over single-link channels\cite{ref:multicast_streaming_it15, ref:layered_it15}. In these works, the channel is adversarially permitted to erase symbols in each channel packet up to a maximum number of erasures within each sliding window. The Rank-Deficient Sliding Window Network can be viewed as an extension of this model, where the channel introduces rank deficiencies rather than erasures.
\end{rem} 

\subsection{Decoder}
Let $T$ be the maximum delay permissible by the receiver node. A packet received at time $t$ must be recovered by time $t+T$ using a delay-constrained decoder, i.e., $ \hat{\bs}_t = \eta_t(\by_0,\dots,\by_{t+T})$ is the reconstructed packet.
If the decoded source packet $\hat{\textbf{s}}_t$ is equal to $\bs_t$, then the source packet is perfectly recovered by the deadline; otherwise, it is declared lost. 
A linear code $\mathcal{C}$ over $\field{F}_{q^M}$ is defined as \textit{feasible} for $\isolatedchannel(S,W)$ if the encoding and decoding functions for the code are capable of perfectly recovering every source packet transmitted over the channel with delay $T$. 

In this paper we will assume that the window length satisfies ${W=T+1}$. A source packet $\bs_t$ must be decoded by time $t+T$ at the receiver. Thus its active duration spans the interval $[t,t+T]$, which maps to a window length of $W=T+1$.  Nevertheless, we will also discuss how our codes can handle the case when $W \neq T+1$. 

Our objective in this paper is to construct codes that guarantee recoverability under the worst channel conditions for a fixed delay and rate, i.e., identifying the largest rank deficiency $S$ for which a code with a given rate is feasible. Towards this end, we introduce a new metric called the column sum rank distance of a convolutional code. We show that it is both a necessary and sufficient metric to determine the maximum rank deficiency from which the code guarantees perfect recovery. Thus, maximizing $S$ reduces to finding codes with maximum column sum rank distance over the interval $[0,T]$. 

Codes which achieve the maximum column sum rank distance will be referred to as Maximum Sum Rank (MSR) codes in this paper. Furthermore the column sum rank distance  possess a profile property. Achieving the maximum  distance at one point implies that it is also maximized at all points before it. Operationally we show that this property guarantees that the product of the generator matrix with elements of a specific set of channel matrices is full-rank. Finally we propose a family of super-regular matrices that permit this multiplication property of the generator matrix, which we then use to construct MSR codes. Our proposed family of codes uses the properties of rank metric block codes and $m$-MDS convolutional codes, which are introduced as preliminaries in the following section.


\section{Background}
\label{sec:prior_codes}

\subsection{Rank Metric Codes}
\label{subsec:rank_metric_codes}

Consider a vector $\bx \in \field{F}_{q^M}^n$ over the extension field. We refer to $\bx$ as a channel packet. The vector $\bx$ over the extension field is isomorphic to an $n \times M$ matrix over the ground field $\field{F}_q$.
Formally, a bijective mapping $\phi_n:\field{F}_{q^M}^n \rightarrow \field{F}_q^{n\times M}$ allows for the conversion of this vector to a matrix over the ground field. We more thoroughly detail this mapping in Appendix \ref{app:background}. Noting that a normal basis can describe every element in the extension field, the elements of $\field{F}_{q^M}$ are mapped to linearized polynomials evaluated at a normal element, whose coefficients form column vectors. A row vector in $\field{F}_{q^M}^n$ then maps to a matrix whose columns are the coefficients of the corresponding linearized polynomials.

The rank of $\bx$ is defined as the rank of its associated matrix $\phi_n(\bx)$. The \textit{rank distance} between any two vectors $\bx,\hat{\bx} \in \field{F}_{q^M}^n$ is defined as
\begin{align*}
d_R(\bx,\hat{\bx}) \triangleq \rank{(\phi_n(\bx) - \phi_n(\hat{\bx}))}.
\end{align*}
The rank distance is a metric and is upper bounded by the Hamming distance \cite{ref:mrd_pit85}. For any linear block code $\mathcal{C}[n,k]$ over $\field{F}_{q^M}$, the \textit{minimum rank distance} is defined as the smallest rank amongst all non-zero channel packets. Similar to the minimum Hamming distance, the minimum rank distance of a code must satisfy a Singleton-like bound, i.e., $d_R(\mathcal{C}) \leq \min \left\{ 1 , \frac{M}{n} \right\} (n-k) + 1$ \cite{ref:mrd_pit85}. We simplify the notation when $\mathcal{C}$ is obvious. It is assumed that $M \geq n$ from here on; $d_R$ is then bounded exactly by the classic Singleton bound. Any code that meets this bound with equality is referred to as a Maximum Rank Distance (MRD) code. Such codes possess the following property.
\begin{thm}[Gabidulin, \cite{ref:mrd_pit85}]
\label{thm:matrix_mult_property} 
Let $\bG \in \field{F}_{q^M}^{k\times n}$ be the generator matrix of an MRD code. The product of $\bG$ with any full-rank matrix $\bA \in \field{F}_q^{n\times k}$ satisfies $\rank{\bG\bA} = k$.
\end{thm}

A complementary theorem was proven in \cite{ref:mrd_pit85} for the parity-check matrix of an MRD code. We use the equivalent generator matrix property, which arises from the fact that the dual of an MRD code is also an MRD code \cite{ref:mrd_pit85}.

Gabidulin codes are an important family of MRD codes. To construct a Gabidulin code, let $g_0,\dots,g_{n-1} \in \field{F}_{q^M}$ be a set of elements that are linearly independent over $\field{F}_q$. The generator matrix for a Gabidulin code $\mathcal{C}[n,k]$ is given by
\begin{align*}
\bG = 
\begin{pmatrix}
g_0 & g_1 & \dots & g_{n-1} \\
g_0^{[1]} & g_1^{[1]} & \dots & g_{n-1}^{[1]} \\
\vdots & \vdots & \ddots & \vdots \\
g_0^{[k-1]} & g_1^{[k-1]} & \dots & g_{n-1}^{[k-1]}
\end{pmatrix}.
\end{align*}
where we use the notation $g^{[j]} \defeq g^{q^j}$ to denote the $j$-th Frobenius power of $g \in \field{F}_{q^M}$ (see Appendix~\ref{app:background}).
Gabidulin codes can be applied directly as end-to-end codes over networks where a channel matrix $\bA \in \field{F}_q^{n\times n}$ transforms symbols of the transmitted channel packet \cite{ref:silva_it08, ref:khisti_isit10}. 
For a source packet $\bs \in \field{F}_{q^M}^k$ encoded by a Gabidulin code, a receiver observes $\by = \bs \bG \bA$. By Theorem \ref{thm:matrix_mult_property}, the product $\bG \bA$ is an invertible matrix as long as $\rank{\bA} \geq k$.

\subsection{The Column Hamming Distance}
\label{subsec:bg_3}

Let $\mathcal{C}[n,k,m]$ be a linear time-invariant convolutional code, where $m$ is the code memory. For a source packet sequence $\bs_{[0,j]} = (\bs_0,\dots,\bs_j) \in \field{F}_{q^M}^{k(j+1)}$, the channel packet sequence\footnote{In network coding literature, each $\bx_t \in \field{F}_{q^M}^n$ is referred to as a generation of $n$ channel packets. We denote $\bx_t$ as a channel packet containing $n$ symbols and $\bx_{[t,t+j]}$ as a sequence of $j$ packets. Similar notation is used for the source.} $\bx_{[0,j]} = \bs_{[0,j]} \bGEX_j$ is determined using the extended form generator matrix
\begin{align}
\label{eq:cc_generator}
\bGEX_j \triangleq \begin{pmatrix}
\bG_0 & \bG_1 & \dots & \bG_j \\
	& \bG_0 & \dots & \bG_{j-1} \\
	&	& \ddots & \vdots \\
	&	&	& \bG_0
\end{pmatrix},
\end{align}
where $\bG_j \in \field{F}_{q^M}^{k \times n}$ and $\bG_j = \mathbf{0}$ for $j > m$ \cite[Chapter~1]{ref:zigangirov_fundamentals99}. 
We assume from here on that $\bG_0$ always has full row rank. This guarantees that $\bGEX_j$ also possesses full row rank, which is a property used in subsequent results.

The Hamming weight of $\bx_{[0,j]}$ is a sum of the Hamming weight of each channel packet
$\bx_t$ for $0\leq t\leq j$. 
The $j$-th column Hamming distance of a convolutional code is defined
\begin{align*}
d_H(j) \triangleq \min_{\bx_{[0,j]} \in \mathcal{C}, \bs_0 \neq \bzero} \wt_H(\bx_{[0,j]}),
\end{align*}
 as the minimum Hamming weight amongst all channel packet sequences for which the initial source packet $\bs_0$ is non-zero \cite{ref:lin_ecc, ref:smds_it06}. Note that because $\bG_0$ is full-rank, $\bs_0 \neq \bzero$ immediately implies that $\bx_0 \neq \bzero$ as well. 

Several properties pertaining to the column Hamming distance were treated in \cite{ref:smds_it06,ref:tomas_it12}. We summarize two relevant ones below. In Section \ref{sec:msr_thms}, we prove analogous properties to these for the rank metric.

\begin{prop}[Tomas et al., \cite{ref:tomas_it12}]
\label{prop:mmds_prop1}
Consider an erasure channel being used for each $t\geq 0$, where the prior source sequence $\bs_{[0,t-1]}$ is known to the decoder by time $t+j$. If there are at most $d_H(j)-1$ symbol erasures in the window $[t,t+j]$, then $\bs_t$ is recoverable by time $t+j$. Conversely, there is at least one hypothetical channel window $[t,t+j]$ containing $d_H(j)$ erasures for which $\bs_t$ is not recoverable by time $t+j$.
\end{prop}
Property \ref{prop:mmds_prop1} states that for sliding window erasure channels featuring windows of length $W$, a convolutional code with column Hamming distance $d_H(W-1)$ can guarantee perfect decoding with delay $W-1$, provided that there are less than $d_H(W-1)$ erasures in the window \cite{ref:layered_it15}.

\begin{prop}[Gluessing-Luerssen et al., \cite{ref:smds_it06}]
\label{prop:mmds_prop2}
The $j$-th column Hamming distance of a code is upper bounded by a Singleton-like bound, i.e., $d_H(j) \leq (n-k)(j+1)+1$. If $d_H(j)$ meets this bound with equality, then $d_H(i)$ meets its respective bound for all $i \leq j$.
\end{prop}
This property asserts a convolutional code extension of the Singleton bound. The desirability of achieving large column Hamming distances is given in the previous Property \ref{prop:mmds_prop1}. In conjunction with this, a code capable of recovering from $d_H(j)-1$ erasures with delay $j$ can be shown to be further capable of the respective maximum recovery for all $i \leq j$.


There exist several families of codes which achieve the maximum $d_H(j)$ for some given $j$ \cite{ref:gabidulin_acct88, ref:smds_it06, ref:hutch_laa08}. One such class of codes are $m$-MDS codes. These  achieve the upper bound up to the code memory i.e., $d_H(j) =(n-k)(j+1)+1$ for $0\le j \le m$. The MSR codes, which we introduce in this work, are rank metric analogues of $m$-MDS codes. One approach to constructing the generator matrix of $m$-MDS codes is by taking a sub-matrix of $k(m+1)$ rows from a block Toeplitz super-regular matrix \cite{ref:smds_it06}. A prior construction of a block Toeplitz super-regular matrix was given in \cite{ref:almeida_laa13}. As this construction is modified for our purposes, we include a summary of this construction in Section~\ref{sec:preservation}, as well as a review of super-regular matrices in Appendix \ref{app:background}. 

\subsection{The Active Column Sum Rank Distance}
\label{subsec:acsr}

Let $\mathcal{C}[n,k,m]$ be a linear time-invariant convolutional code over $\field{F}_{q^M}$, whose codewords are generated using \eqref{eq:cc_generator}. 
 The \textit{active column sum rank distance} is a metric for convolutional codes that was proposed in a prior work\cite{ref:antonia_14}. This metric is defined using the state Trellis graph of the convolutional code. Let $\mathcal{C}^{a}_j$ be the set of all channel packet sequences $\bx_{[0,j]}$ that are constructed by exiting the zero-state of the Trellis at time $0$ and not re-entering it for $1 \leq t \leq j-1$. The $j$-th active column sum rank of a linear convolutional code $\mathcal{C}[n,k,m]$ is then defined as as the minimum sum rank of all channel packet sequences in $\mathcal{C}^{a}_j$, i.e.,
\begin{align*}
d^a_R(j) & \triangleq \min_{\bx_{[0,j]} \in \mathcal{C}^{a}_j} \sum_{t=0}^j \rank{(\phi_n(\bx_t))},
\end{align*}
where $\phi_n(\cdot)$ is the previously introduced mapping from vectors in the extension field to matrices in the ground field in Section~\ref{subsec:rank_metric_codes}.

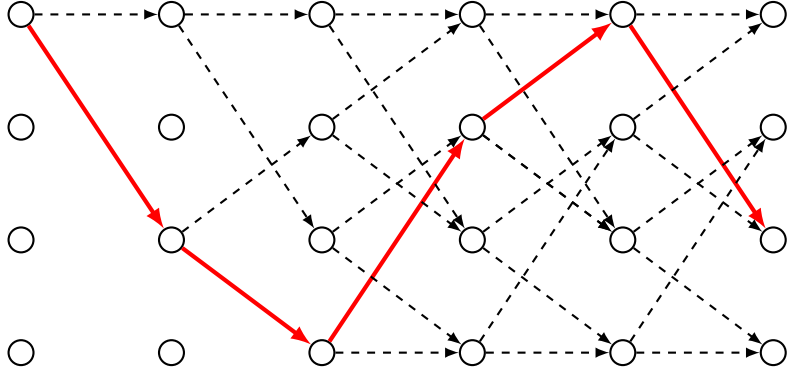
\begin{figure*}
\centering
\begin{tikzpicture}

\foreach \i in {1,...,4} {
	\foreach \j in {1,...,6} {
		\node[draw,circle,thick] (\i\j) at (-5+2*\j,-1.5*\i) {};
	}
}

\draw [-latex,ultra thick, red](11) -- (32);
\draw [-latex,ultra thick, red](32) -- (43);
\draw [-latex,ultra thick, red](43) -- (24);
\draw [-latex,ultra thick, red](24) -- (15);
\draw [-latex,ultra thick, red](15) -- (36);

\draw [-latex,thick, dashed](12) -- (33);
\draw [-latex,thick, dashed](13) -- (34);
\draw [-latex,thick, dashed](14) -- (35);
\draw [-latex,thick, dashed](11) -- (12);
\draw [-latex,thick, dashed](12) -- (13);
\draw [-latex,thick, dashed](13) -- (14);
\draw [-latex,thick, dashed](14) -- (15);
\draw [-latex,thick, dashed](15) -- (16);

\draw [-latex,thick, dashed](23) -- (14);
\draw [-latex,thick, dashed](23) -- (34);
\draw [-latex,thick, dashed](24) -- (35);
\draw [-latex,thick, dashed](24) -- (35);
\draw [-latex,thick, dashed](25) -- (16);
\draw [-latex,thick, dashed](25) -- (36);

\draw [-latex,thick, dashed](33) -- (44);
\draw [-latex,thick, dashed](34) -- (45);
\draw [-latex,thick, dashed](32) -- (23);
\draw [-latex,thick, dashed](33) -- (24);
\draw [-latex,thick, dashed](34) -- (25);
\draw [-latex,thick, dashed](35) -- (26);
\draw [-latex,thick, dashed](35) -- (46);

\draw [-latex,thick, dashed](44) -- (25);
\draw [-latex,thick, dashed](43) -- (44);
\draw [-latex,thick, dashed](44) -- (45);
\draw [-latex,thick, dashed](45) -- (46);
\draw [-latex,thick, dashed](45) -- (26);

\end{tikzpicture}
\caption{A hypothetical transition along the state Trellis graph is highlighted. The active column sum rank distance $d^a_R(5)$ does not guarantee the sum rank of the channel packet sequence generated by this path, i.e., $\bx_{[0,5]} \notin \mathcal{C}^{a}_5$. In this work, we introduce the column sum rank, which does consider this channel packet sequence.}
\label{fig:state_trellis_diagram}
\end{figure*}

Note that by restricting itself to only consider channel packet sequences in $\mathcal{C}^{a}_j$, the active column sum rank of a convolutional code does not impose any guarantees on the sum rank of any channel packets that enter the zero state before time $j$. An example of such a sequence is provided in Fig. \ref{fig:state_trellis_diagram}. Consequently, the active column sum rank is not a sufficient metric to guarantee delay-constrained decoding over a sliding window channel, as we show in the next section.

\section{The Column Sum Rank Distance}
\label{sec:msr_thms}

Let $\mathcal{C}[n,k,m]$ be a linear time-invariant convolutional code over $\field{F}_{q^M}$, constructed in the same manner as in the previous section.
We introduce the $j$-th column sum rank distance of a code 
\begin{align*}
d_R(j) & \triangleq \min_{\bx_{[0,j]} \in \mathcal{C}, \bs_0 \neq 0} \sum_{t=0}^j \rank{(\phi_n(\bx_t))},
\end{align*}
as an analogue of the column Hamming distance. 
Unlike the $j$-th active column sum rank distance, this metric permits returning to the zero-state before time $j$. For example, the channel packet sequence generated in Fig. \ref{fig:state_trellis_diagram} is valid for the column sum rank distance. As a result, this metric is stronger than the active version, i.e., $d_R(j) \leq d^a_R(j)$. In the following theorem, we show that the column sum rank distance of a convolutional code is both necessary and sufficient to guarantee low-delay decoding over $\isolatedchannel(S,W)$. 


\begin{thm}
\label{thm:sum_rank_decoding}
Let $\mathcal{C}[n,k,m]$ be a convolutional code used over the window $[0,W-1]$. For $0 \leq t \leq W-1$, let $\bA^*_t \in \field{F}_q^{n\times \rho_t}$ be full-rank matrices and $\bA_{[0,W-1]}^* = \diag{(\bA_0^*,\dots,\bA_{W-1}^*)}$ be a channel matrix. The following statements are true:
\begin{enumerate}
\item If $d_{R}(W-1) > nW - \sum_{t=0}^{W-1} \rho_t$, then $\bs_0$ is always recoverable by time $W-1$.

\item If $d_{R}(W-1) \leq nW - \sum_{t=0}^{W-1} \rho_t$, then there exists at least one channel packet sequence and channel matrix for which $\bs_0$ is not recoverable by time $W-1$.
\end{enumerate}
\end{thm}

\begin{IEEEproof}
Due to code linearity, we only need to show that all output channel packet sequences are distinguishable from the all-zero sequence. We prove this by contradiction. Consider a source packet sequence $\bs_{[0,W-1]} = (\bs_0,\dots,\bs_{W-1})$, where $\bs_0 \neq \bzero$. Suppose that this sequence generates the channel packet sequence $\bx_{[0,W-1]}$, for which $\bx_{[0,W-1]} \bA_{[0,W-1]}^* = \bzero$. This implies that $\rank{(\phi_n(\bx_t))} \leq n - \rank{(\bA_t^*)}$ for $0\leq t \leq W-1$. By summing each of the inequalities, we arrive at the following contradiction on the sum rank of the channel packet sequence:
\begin{align*}
\sum_{t=0}^{W-1} \rank{(\phi_n(\bx_t))} 	&\leq nW - \sum_{t=0}^{W-1}\rho_t \\
											& < d_R(W-1).
\end{align*}

For the converse, let $\bs_{[0,W-1]} = (\bs_0,\dots,\bs_{W-1})$, with $\bs_0 \neq \bzero$ be a source packet sequence that maps to $\bx_{[0,W-1]}$, for which $\sum_{t=0}^{W-1} \rank{(\phi_n(\bx_t))} = d_R(W-1)$. For $0 \leq t \leq W-1$, let $\rho_t = n - \rank{(\phi_n(\bx_t))}$. There exist matrices $\bA_t^* \in \field{F}_q^{n\times \rho_t}$ such that each $\bx_t \bA_t^* = \bzero$. We let $\bA_{[0,W-1]}^* = \diag{(\bA_0^*,\dots,\bA_{W-1}^*)}$ be the channel matrix. Summing all of the $\rho_t$ reveals $\rank{(\bA_{[0,W-1]}^*)} = nW - d_R(W-1)$. Furthermore, $\bs_{[0,W-1]}$ is indistinguishable from the all-zero source packet sequence over this channel.
\end{IEEEproof}

\begin{rem}
The constraint that $\bs_0 \neq \bzero$ is necessary in order to differentiate the first source packet from the all-zero source packet. Note however, that there is no necessary constraint on the state Trellis transitions; any non-zero source packet sequence should be differentiable from the all-zero sequence. Using the active column sum rank in place of the column sum rank in the above Theorem then leads to a partial guarantee. If $\bx_{[0,W-1]} \in \mathcal{C}^{a}_{W-1}$, then the active column sum rank determines the maximum rank deficiency in the channel from which the first source packet is recoverable. However, if $\bx_{[0,W-1]} \notin \mathcal{C}^{a}_{W-1}$, then the $W-1$-th active column sum rank does not provide any guarantees on recoverability. Consequently, we view the active column sum rank as an over-estimate of the recovery capability of the code.
\end{rem}

For time-invariant encoders, Theorem \ref{thm:sum_rank_decoding} can be used to guarantee that all source packets are recovered with delay at most $W-1$ over a sliding window channel. Assuming all prior packets have been decoded, we recover each $\bs_t$ using the window $[t,t+W-1]$. The contributions of $\bs_{[0,t-1]}$ can be negated from the received packet sequence used for decoding. Theorem \ref{thm:sum_rank_decoding} is then effectively a rank metric analogue to Property \ref{prop:mmds_prop1} from Section \ref{subsec:bg_3}, which describes how column Hamming distance bounds the number of tolerable erasures in single-link streaming \cite{ref:tomas_it12}.

\begin{rem}
Aside from rank-deficient channel matrices, adversarial errors can also be considered using rank metric codes. Consider a single-link single-shot system where the receiver observes $\by = \bx + \be$, with $\be \in \field{F}_{q^M}^n$ being an additive error vector. If $\rank{(\phi_n(\be))} \leq \frac{d_R - 1}{2}$, the decoder for an MRD code can recover the source \cite{ref:mrd_pit85}. MRD codes reflect a rank analogue of the error correcting capability of MDS codes. It can easily be shown that the column sum rank can ensure $\bx_0$ is recoverable by time $j$ in the channel $\by_{[0,j]} = \bx_{[0,j]} + \be_{[0,j]}$, if the sum rank of $\be_{[0,j]} \in \field{F}_{q^M}^{n(j+1)}$ is constrained to be at most $\frac{d_R(j) - 1}{2}$. The proof follows similarly to that for Theorem \ref{thm:sum_rank_decoding}.
\end{rem}

We next propose an analogue to Property \ref{prop:mmds_prop2} from Section \ref{subsec:bg_3}. First, we bound the column sum rank by $d_R(j) \leq (n-k)(j + 1) + 1$. The sum rank of a channel packet cannot exceed its Hamming weight, meaning that the upper bound on column Hamming distance is inherited by the rank metric analogue.  Furthermore, we show that if the $j$-th column sum rank achieves its upper bound, all prior column sum ranks do so as well for their respective bounds.

\begin{lem}
\label{lem:sum_rank_profile}
If $d_R(j) = (n-k)(j+1)+1$, then $d_R(i) = (n-k)(i+1)+1$ for all $i \leq j$.
\end{lem}

\begin{IEEEproof}
It suffices to prove for $i = j-1$. Let $\mathcal{C}[n,k,m]$ be a code for which $d_R(j-1)$ does not attain the upper bound i.e.,  $d_R(j-1) \le (n-k)j$, but $d_R(j)$ achieves the maximum i.e., $d_R(j) = (n-k)(j+1)+1$. We will argue by contradiction that such a code cannot exist. 

Consider a source packet sequence $\bs_{[0,j-1]}$ that generates $\bx_{[0,j-1]}$ whose sum rank equal to $d_R(j-1)$ i.e., $\sum_{t=0}^{j-1}\rank{(\phi_n(\bx_t))} = d_R(j-1)$ holds. We argue that this sequence can be augmented to include $\bx_j$, such that $\sum_{t=0}^j \rank{(\phi_n(\bx_t))} \le (n-k)(j+1) < d_R(j) $ holds. This will complete the contradiction.

To exhibit such a choice of $\bx_j$ recall that $\bx_j = \sum_{t = 0}^{j-1} \bs_t \bG_{j-t} + \bs_j \bG_0$. The summation up to $j-1$ produces a vector whose Hamming weight is at most $n$. Because $\rank{(\bG_0)} = k$, the source packet $\bs_j$ can be selected specifically in order to negate up to $k$ non-zero entries of the first summation. This implies that $\wt_H{(\bx_j)} \leq n-k$ and consequently, $\rank{(\phi_n(\bx_j))} \leq n-k$. 
Therefore, we bound the sum rank of $\bx_{[0,j]}$
\begin{align*}\sum_{t=0}^j \rank{(\phi_n(\bx_t))} &= d_R(j-1) + \rank{(\phi_n(\bx_j))}\\ & \leq d_R(j-1) + n-k \\
	& \leq (n-k)(j+1),
\end{align*} 
as required. 
\end{IEEEproof}

Codes achieving the Singleton bound for $d_R(m)$ are referred to as MSR codes. They directly parallel $m$-MDS codes, which maximize the $m$-th column Hamming distance \cite{ref:smds_it06}. In fact, since $d_R(j) \leq d_H(j)$, MSR codes automatically maximize the column Hamming distance and can be seen as a special case of $m$-MDS codes. 

By Theorem \ref{thm:sum_rank_decoding}, an MSR code with memory $T = W-1$ recovers source packets with delay $T$, when the rank of the channel matrix is at least $k(T+1)$ in each sliding window of length $W$. We prove the existence of MSR codes in the next section, but first discuss a matrix multiplication property for the generator matrix. 
The following theorem serves as an extension of Theorem \ref{thm:matrix_mult_property} to convolutional codes transmitted over independent network uses.

\begin{thm}
\label{thm:sum_rank_construct}
For $0 \leq t \leq j$, let $0 \leq \rho_t \leq n$ satisfy
\begin{align}
\label{eq:thm_sum_rank_constraint}
\sum_{i=0}^t \rho_i & \leq k(t+1)
\end{align}
for all $t \leq j$ and with equality for $t = j$. The following are equivalent for any convolutional code:

\begin{enumerate}
\item $d_R(j) = (n-k)(j+1)+1$
\item For all full-rank $\bA_{[0,j]}^* = \diag{(\bA_0^*,\dots,\bA_j^*)}$ constructed from full-rank blocks $\bA_t^* \in \field{F}_q^{n\times \rho_t}$ and $\rho_t$ that satisfy \eqref{eq:thm_sum_rank_constraint}, the product $\bGEX_j \bA_{[0,j]}^*$ is non-singular. 
\end{enumerate}
\end{thm}

\begin{IEEEproof}We first prove $1 \Rightarrow 2$.  Suppose there exists an $\bA_{[0,j]}^*$ whose blocks satisfy \eqref{eq:thm_sum_rank_constraint}, for which $\bGEX_j \bA_{[0,j]}^*$ is singular. 
Then there exists a channel packet sequence $\bx_{[0,j]}$, where $\bx_{[0,j]} \bA_{[0,j]}^* = \bzero$. We show that this leads to a contradiction of 1. The contradiction is immediate if $\bx_0 \neq \bzero$. In this case the sum rank of $\bx_{[0,j]}$ is at least $d_R(j)$, i.e., 
$\sum_{t=0}^j \rank{(\phi_n(\bx_t))} \ge d_R(j)$ must hold. Note however that:
\begin{align*}
\sum_{t=0}^j \rank{(\phi_n(\bx_t))} & \leq n(j+1) - \sum_{t=0}^j \rho_t \\
& = (n-k)(j+1)
\end{align*}
contradicts $d_R(j) = (n-k)(j+1)+1$. Note that we use~\eqref{eq:thm_sum_rank_constraint} with equality at $t=j$ in the second step. 

If $\bx_0 = \bzero$, then the sum rank of $\bx_{[0,j]}$ is not constrained by $d_R(j)$.  Let $l = \argmin_{t} \bx_t \neq \bzero$ be the smallest index for which $\bx_l$ is non-zero and consider the channel packet sequence $\bx_{[l,j]}$, whose sum rank is at least $d_R(j-l)$. Because $\bx_t \bA_t^* = \bzero$ for $t = l,\dots,j$, we bound $\rank{(\phi_n(\bx_t))} \leq n - \rho_t$ in this window. The sum rank of $\bx_{[l,j]}$ is bounded:
\begin{align}
\sum_{t=l}^j \rank{(\phi_n(\bx_t))} & \leq n(j-l+1) - \sum_{t=l}^j \rho_t \nonumber\\
& \leq (n-k)(j-l+1). \nonumber
\end{align}
The second line follows from $\sum_{t=l}^j \rho_t \geq k(j-l+1)$, which can be derived when \eqref{eq:thm_sum_rank_constraint} is met with equality for $t=j$. Due to Lemma \ref{lem:sum_rank_profile}, the column sum rank achieves $d_R(j-l) = (n-k)(j-l+1)+1$. The sum rank of $\bx_{[l,j]}$ is less than $d_R(j-l)$, which is a contradiction.

We prove $2 \Rightarrow 1$ by using a code with $d_R(j) \leq (n-k)(j+1)$ and constructing a full-rank $\bA_{[0,j]}^*$ for which $\bGEX_j \bA_{[0,j]}^*$ is singular. Let $m=\argmin_i d_R(i) \leq (n-k)(i+1)$ be the first instance where the column sum rank fails to attain its upper bound and consider the sequence $\bx_{[0,m]}$ with the minimum column sum rank. We show that there exist full-rank matrices $\bA_t^* \in \field{F}_q^{n\times \rho_t}$ satisfying both \eqref{eq:thm_sum_rank_constraint} and  $\bx_t \bA_t^* = \bzero$ for $0 \leq t \leq m$. In addition, we aim to have equality in~\eqref{eq:thm_sum_rank_constraint} at $t=m$ and thus  $\bA_{[0,m]}^*$ will be of dimension $(m+1)n\times (m+1)k$. This is relevant later in the proof.

When $m=0$, the column rank of $\bx_0$ cannot exceed $n-k$. For every $\rho_0 \leq n - \rank{(\phi_n(\bx_0))}$, there exists an $\bA_0^*$ for which $\bx_0 \bA_0^* = \bzero$. Clearly we can always choose an $\bA_0^*$ with rank $\rho_0 = k$.

When $m>0$, the sum rank of $\bx_{[0,t]}$ satisfies
\begin{align}
\label{eq:cond_t_1}
\sum_{i=0}^t \rank(\phi_n(\bx_i)) \ge (n-k)(t+1)+1
\end{align}
for $0 \le t \le m-1$, and
\begin{align}
\label{eq:cond_t_2}
\sum_{i=0}^m \rank(\phi_n(\bx_i)) \le (n-k) (m+1). 
\end{align} 
Let $\rho_t = n - \rank{(\phi_n(\bx_t))}$ for $0\leq t \leq m-1$ and choose the appropriate $\bA^*_t$ for which $\bx_t \bA^*_t = \bzero$. For all $0 \le t \le m-1$, we have that:
\begin{align}
\sum_{i=0}^t \rho_i & = n(t+1) - \sum_{i=0}^t \rank{(\phi_n(\bx_i))} \label{eq:rho_expr}\\
& \leq k(t+1) - 1, \label{eq:rho_expr2}
\end{align}
confirming that \eqref{eq:thm_sum_rank_constraint} is satisfied for $t \leq m-1$.
Note that in~\eqref{eq:rho_expr2} we apply the inequality in~\eqref{eq:cond_t_1}.  We next specify an appropriate choice for $\rho_m$. We will select:
\begin{align}
\rho_m = (m+1)k - \sum_{t=0}^{m-1} \rho_t, \label{eq:rho_m_select}
\end{align}
and show that there exists an associated $\bA^*_m \in {\mathbb F}_q^{n \times \rho_m}$ that will satisfy $\bx_m \bA^*_m = {\bf 0}$. It thus suffices to show that $\rho_m$ also satisfies   $\rho_m \le n - \rank(\phi_n(\bx_m))$. Note that
\begin{align}
n-\rho_m &= n - (m+1)k + \sum_{i=0}^{m-1} \rho_i \\
&= (n-k)(m+1) -  \sum_{i=0}^{m-1} \rank{(\phi_n(\bx_i))} \label{eq:rho_sub} \\
&=(n-k)(m+1) -  \sum_{i=0}^{m} \rank{(\phi_n(\bx_i))}  \nonumber \\& \quad+ \rank{(\phi_n(\bx_m))} \\
&\ge \rank{(\phi_n(\bx_m))}, \label{eq:lb_xm}
\end{align}
where~\eqref{eq:rho_sub} follows by using~\eqref{eq:rho_expr} with ${t=m-1}$ and~\eqref{eq:lb_xm} follows via the inequality in~\eqref{eq:cond_t_2}.
Finally, our choice~\eqref{eq:rho_m_select} also guarantees that $\bA_{[0,m]}^*$ has dimension $(m+1)n \times (m+1)k$ as claimed.

The remaining $\bA_{m+1}^*,\dots,\bA_j^*$ can be any full-rank $n\times k$ matrices, thus satisfying \eqref{eq:thm_sum_rank_constraint} for all $t \leq j$. The product $\bGEX_j \bA^*_{[0,j]}$ can be written as
\begin{align*}
\bGEX_j \bA^*_{[0,j]} &=
\begin{pmatrix}
\bGEX_m & \bX \\
	& \bY
\end{pmatrix}
\begin{pmatrix}
\bA_{[0,m]}^* \\
	& \bA_{[m+1,j]}^*
\end{pmatrix} \\
&=
\begin{pmatrix}
\bGEX_m \bA_{[0,m]}^* & \bX \bA_{[m+1,j]}^* \\
	& \bY \bA_{[m+1,j]}^*
\end{pmatrix}
\end{align*}
where $\bX$ and $\bY$ denote the remaining blocks that comprise $\bGEX_j$. The block $\bGEX_m \bA_{[0,m]}^*$ is a square matrix with a zero determinant. Therefore, $\det{\bGEX_j \bA_{[0,j]}^*}$ is also zero.
\end{IEEEproof}


Although in this work, we do not propose a general decoding algorithm for MSR codes, we remark that decoding in the network streaming problem can be reduced to matrix inversion. Consider a scenario where an MSR code with memory $T=W-1$ is used and that all source packets before time $t$ have been recovered. To ensure that $\bs_t$ is recoverable within its deadline of time $t+T$, we consider increasingly larger windows $[t,t+j]$ for $0 \leq j \leq T$. Theorem \ref{thm:sum_rank_decoding} states that if $\sum_{i=t}^{t+j} \rho_i < k(t+j+1)$, then the decoder cannot guarantee recovery by time $t+j$. The window length $j$ is incremented at each time instance up to the first point where $\sum_{i=t}^{t+j} \rho_i \geq k(t+j+1)$ is achieved. By Theorem \ref{thm:sum_rank_construct}, the rank conditions in \eqref{eq:thm_sum_rank_constraint} are satisfied and $\bGEX_j \bA^*_{[t,t+j]}$ is non-singular. We therefore invert this matrix and solve for $\bs_{[t,t+j]}$, recovering all packets in the window simultaneously. Consequently, packets encoded by an MSR code can be recovered for the network streaming problem with complexity $\mathcal{O}((jk)^3)$.

In the next section, we construct an extended generator matrix. Theorem \ref{thm:sum_rank_construct} is then useful afterwards to verify that the generator matrix does in fact define an MSR code.

\section{Preservation of Super-regularity}
\label{sec:preservation}

In \cite{ref:smds_it06}, the authors provided a construction for a Toeplitz super-regular matrix that exists in $\field{F}_q$ for a prime $q$. 
From \eqref{eq:cc_generator}, the generator matrix of a convolutional code is block Toeplitz and we focus on super-regular matrices with similar structure. For simplicity, we consider block Hankel matrices and later convert the structure to block Toeplitz before code construction.

\subsection{Block Hankel Super-regular Matrices}

A block Hankel super-regular matrix construction, which we outline below, was proposed in \cite{ref:almeida_laa13} for $\field{F}_{q^M}$ where $q$ is a prime power and $M$ is sufficiently large.

\begin{thm}[Almeida et al., \cite{ref:almeida_laa13}]
	For $n,m \in \field{N}$, let $M=q^{n(m+2)-1}$. Let $\alpha \in \field{F}_{q^M}$ be a primitive element and a root of the minimal polynomial $p_\alpha(X)$. For $0 \leq j \leq m$, let the blocks $\bT_j \in \field{F}^{n\times n}_{q^M}$ be defined by
	\begin{align}
	\label{eq:sr_const}
	\bT_j =
	\begin{pmatrix}
	\alpha^{[nj]} & \alpha^{[nj+1]} & \dots & \alpha^{[n(j+1)-1]} \\
	\alpha^{[nj+1]} & \alpha^{[nj+2]} & \dots & \alpha^{[n(j+1)]} \\
	\vdots & \vdots & \ddots & \vdots \\
	\alpha^{[n(j+1)-1]} & \alpha^{[n(j+1)]} & \dots & \alpha^{[n(j+2)-2]} \\
	\end{pmatrix}.
	\end{align}
Then the following block Hankel matrix
	\begin{align}
	\label{eq:toeplitz_esr_rev}
	{\bT} = 
	\begin{pmatrix}
	&	&	&	 \bT_0 \\
	&	& \bT_0 & \bT_1 \\
	& \iddots & \vdots & \vdots \\
	\bT_0 &  \dots & \bT_{m-1} & \bT_m
	\end{pmatrix}
	\end{align}
is super-regular.
\end{thm}

We refer the reader to~\cite{ref:almeida_laa13} for the full proof. In this subsection however, several properties of $\bT$ that lead to its super-regularity are highlighted. They will be useful in the subsequent section. 

Let ${T}_{j}(r,s)$ be the element in the $r$-th row and $s$-th column of ${\bT}_j$, where $0 \leq r,s \leq n-1$ and $0 \leq j \leq m$. 
Each entry of the matrix is a linearized monomial evaluated at $X=\alpha$, in particular
\footnote{We use the variable $X$ when discussing properties of linearized polynomials and evaluate the polynomials at $X=\alpha$ specifically when calculating the determinant of a matrix. However, we suppress the dependence on $X$ whenever it can be determined from the context.},
\begin{align}
{T}_{j}(r,s) = \alpha^{[nj+r+s]}.
\end{align} 
The $q$-degrees of these monomials increase as one moves further down and to the right inside any ${\bT}_j$. 
This can be described with the following inequalities:
\begin{subequations}
	\label{eq:T_deg_ineq_within}
	\begin{align}
	\qdeg{{T}_{j}(r',s)} &> \qdeg{{T}_{j}(r,s)}, \quad 0 \leq r < r' \leq n-1, \label{eq:t_deg_ascend_down1}\\
	\qdeg{{T}_{j}(r,s')} &> \qdeg{{T}_{j}(r,s)}, \quad 0 \leq s < s' \leq n-1, \label{eq:t_deg_ascend_right1}
	\end{align}
\end{subequations}
for all $0 \leq j \leq m$.

Combining all of the $\bT_j$ to construct ${\bT}$ in~\eqref{eq:toeplitz_esr_rev} preserves the property of $q$-degrees strictly increasing downwards (along a single column) or rightwards (along a single row) now for the block Hankel matrix, i.e.,  
\begin{subequations}
	\label{eq:T_deg_ineq_across}
	\begin{align}
	\qdeg{{T}_{j'}(r',s)} &> \qdeg{{T}_{j}(r,s)}, \quad 0 \leq j < j' \leq m, \label{eq:t_deg_ascend_down2}\\
	\qdeg{{T}_{j'}(r,s')} &> \qdeg{{T}_{j}(r,s)}, \quad 0 \leq j < j' \leq m, \label{eq:t_deg_ascend_right2}
	\end{align}
\end{subequations}
for all $0 \leq r, r' \leq n-1$ and $0 \leq s, s' \leq n-1$.


To show that ${\bT}$ is super-regular, let $\bD$ be any $l \times l$ sub-matrix of ${\bT}$. All sub-matrices of ${\bT}$ preserve the degree inequalities in~\eqref{eq:T_deg_ineq_within} and~\eqref{eq:T_deg_ineq_across}. 
${\bT}$ being super-regular is equivalent to $\det{\bD} \neq 0$ for every $\bD$ with a non-trivial determinant.

The determinant is evaluated using the Leibniz formula as a polynomial $D(\alpha) = \sum_{\sigma \in S_r} \sgn{(\sigma)} D_\sigma$.
Each non-zero term in the formula, $D_\sigma$, is a product of linearized monomials
$$D_\sigma = \prod_{i=0}^{l-1} D(i, \sigma(i)).$$ 
Note that the determinant polynomial itself is not linearized. Using~\eqref{eq:T_deg_ineq_within} and~\eqref{eq:T_deg_ineq_across}, the degree of this polynomial can be bounded.
\begin{lem}[Almeida et al., \cite{ref:almeida_laa13}]
	\label{lem:sr_prop}
	For ${\bT}$ defined in \eqref{eq:toeplitz_esr_rev}, let $\bD$ be any square sub-matrix with a non-trivial determinant (see Appendix~\ref{subsec:bg_4}). Let $D(X)$ be the polynomial, such that $D(\alpha) = \det{\bD}$. The degree of $D(X)$ is bounded by
	\begin{align*}
	1 \leq \deg{D(X)} <q^{n(m+2)-1}.
	\end{align*}
\end{lem}

In \cite{ref:almeida_laa13}, the matrices defined by \eqref{eq:sr_const} contained entries with Frobenius powers with fixed $q=2$, rather than for a general $q$. 
A direct extension reveals that the bounds in Lemma~\ref{lem:sr_prop} still holds when using Frobenius powers with arbitrary $q$ in~\eqref{eq:sr_const} provided that $M \geq q^{n(m+2)-1}$.
We refer the reader to \cite{ref:almeida_laa13,ref:mahmood_phd} for the full proof.

An arbitrary $q$ and appropriate choice of $\alpha$ permits the super-regular blocks $\bT_j$ to resemble Gabidulin code generator matrices. 
We require this slight generalization since our channel in Section~\ref{sec:network_streaming_problem} can operate over any field $\field{F}_q$.


The upper bound gives $\deg{D(X)} < \deg{p_\alpha(X)}$, which implies that, $p_\alpha(X) \nmid D(X)$. Consequently, $\alpha$ cannot be a root of $D(X)$ (see Lemma \ref{lem:min_polynomial_divides} of Appendix \ref{app:background}). 


The lower bound in our extension does not change from the original. It implies that $D(X)$ is not the zero polynomial and can be derived by algorithmically finding a unique permutation $\bar{\sigma} = \argmax_{\sigma}{\deg{D_\sigma(X)}}$, which generates the highest degree monomial term in the Leibniz formula. Because $\bar{\sigma}$ is unique, there is no other term which can negate this monomial. Then, $\deg{D(X)} = \deg{D_{\bar{\sigma}}(X)}$, meaning $D(X)$ is not the zero polynomial.

The upper and lower bounds cover both potential cases to ensure $D(\alpha) \neq 0$. Therefore, any $\bD$ with a non-trivial determinant is also non-singular and consequently, ${\bT}$ is a super-regular matrix.

A block Toeplitz or Hankel super-regular matrix can be used to construct the extended generator matrix of an $m$-MDS code~\cite{ref:hutch_laa08}. 
We will use a similar technique to construct MSR codes.
Super-regularity alone however, is insufficient for the network streaming problem. 
We show that the super-regular matrix construction technique above can be modified to produce the desired property for the network streaming problem.


\subsection{Preservation of Super-regularity after Multiplication}

In this section, we connect the matrix multiplication property in Theorem \ref{thm:sum_rank_construct} to super-regularity. The block Hankel matrix $\bT$ from~\eqref{eq:toeplitz_esr_rev} is constructed using an $\alpha$ that is both a primitive and a normal element of the field $\field{F}_{q^M}$. We show in this case that the super-regularity of $\bT$ is preserved after multiplication with any block diagonal matrix in the ground field.

\begin{thm}
	\label{thm:esr}
	For $0 \leq t \leq m$, let $\bA_t \in \field{F}_q^{n\times n}$ be any non-singular matrices. We then construct $\bA_{[0,m]} = \diag{(\bA_0,\dots,\bA_m)}$.
	Let ${\bT} \in \field{F}_{q^M}^{n(m+1) \times n(m+1)}$ be the super-regular matrix in \eqref{eq:toeplitz_esr_rev}.
	If $M \geq q^{n(m+2)-1}$ and $\alpha$ is a primitive normal element of $\field{F}_{q^M}$, then $\bF = {\bT} \bA_{[0,m]}$ is super-regular.
\end{thm}

Several properties of $\bF$ are introduced before proving Theorem~\ref{thm:esr}.
To simplify the notation, we assume that $\bA_0 = \bA_1 = \dots = \bA_m = \bA$.
This allows free use of the polynomial structures and bounds from the previous subsection.
We will then explain why the proof extends immediately to the general case. Hence, $\bA_{[0,m]} = \diag{(\bA,\dots,\bA)}$, where $\bA \in \field{F}_q^{n\times n}$ is a non-singular matrix in the ground field and we consider properties of the product $$\bF = {\bT} \bA_{[0,m]}.$$


\begin{prop}[]
\label{prop:F_block_hankel}
The product $\bF = {\bT} \bA_{[0,m]}$ is a block Hankel matrix, whose blocks each possess a Moore structure.
\end{prop}

The product can be written as
\begin{align*}
\bF = {\bT} \bA_{[0,m]} &= 
\begin{pmatrix}
&	&	&	 \bT_0 \bA \\
&	& \bT_0 \bA & \bT_1 \bA \\
& \iddots & \vdots & \vdots \\
\bT_0 \bA &  \dots & \bT_{m-1} \bA & \bT_m \bA
\end{pmatrix}.
\end{align*}
Let $\bF_j = \bT_j \bA$ for $j = 0,\dots,m$. 
To show that $\bF_j$ has a Moore structure, let $(f_0,\dots,f_{n-1})$ be the first row of $\bF_0$, i.e.,
\begin{align*}
(f_0,\dots,f_{n-1}) = (\alpha^{[0]},\dots,\alpha^{[n-1]}) \bA.
\end{align*}
Each element $f_i$ is a linearized polynomial $f_i(X)$ evaluated at $X=\alpha$, with coefficients from $\bA = [A_{l,i}]$, i.e.,
\begin{align}
\label{eq:lin_pol_prop}
f_i = \sum_{l=0}^{n-1} A_{l,i} \alpha^{[l]}.
\end{align}
Since $\alpha \in \field{F}_{q^M}$ and $A_{l,i} \in \field{F}_q$, invoking Freshman's rule
\begin{align}
\label{eq:freshmans}
f_i(X^{[s]}) = f_i^{[s]}(X), \quad  s \geq 0, 
\end{align}
results in the following Moore structure,
\begin{flalign}
\label{eq:esr_blocks}
\bF_j = 
\begin{pmatrix}
f_0^{[nj]} & f_1^{[nj]} & \dots & f_{n-1}^{[nj]} \\
f_0^{[nj+1]} & f_1^{[nj+1]} & \dots & f_{n-1}^{[nj+1]} \\
\vdots & \vdots & \ddots & \vdots \\
f_0^{[n(j+1)-1]} & f_1^{[n(j+1)-1]} & \dots & f_{n-1}^{[n(j+1)-1]} 
\end{pmatrix}
\end{flalign}
for $0 \leq j \leq m$.

\begin{rem}
	\label{rem:independent_rows}
	Since $\alpha$ is  normal over $\field{F}_{q^M}$ and $\bA$ is full rank,  $(f_0,\dots,f_{n-1})$ are linearly independent over $\field{F}_q$ (see~\cite{ref:mrd_pit85}).
\end{rem}


\begin{prop}
\label{prop:sorted_qdegs}
The $q$-degrees of the polynomial elements of $\bF_j$  strictly increase downwards on any fixed column. They are not necessarily monotonic across a fixed row.
\end{prop}

From~\eqref{eq:lin_pol_prop} and~\eqref{eq:freshmans}, the $q$-degree of each linearized polynomial is bounded by
\begin{align}
\label{eq:f_deg_bounds}
s \leq \qdeg{f_i^{[s]}}(X) \leq n-1+s
\end{align}
for $0 \leq s \leq n(m+1)-1$.
Note that both the upper and lower bounds in \eqref{eq:f_deg_bounds} do not depend on the column index $i$. Furthermore all elements on any fixed row in $\bF_j$ share the same degree (c.f.~\eqref{eq:esr_blocks}).
Consequently, the polynomial entries on any fixed row of the block $\bF_j$ share the same bound and are not necessarily in an increasing order.
Thus we do not have a counterpart of~\eqref{eq:t_deg_ascend_right1} that guarantees that the q-degrees of elements are monotonically increasing across each row in $\bT_j$.

The counterpart of~\eqref{eq:t_deg_ascend_down1} is satisfied however.  Let $F_{j}(r,s)$ be the element in $r$-th row and $s$-th column of $\bF_j$, i.e.,
\begin{align}
\label{eq:Fjrs_def}
F_{j}(r,s) = f_s^{[nj+r]},
\end{align}
 where $0 \leq r,s \leq n-1$ and $0 \leq j \leq m$. The Moore structure of $\bF_j$ implies that the $q$-degrees of the polynomial entries strictly increase from top to bottom along any fixed column of $\bF_j$, i.e.,
\begin{align}
\qdeg{F_{j}(r',s)} > \qdeg{F_{j}(r,s)}, \quad 0 \leq r < r' \leq n-1, \label{eq:f_deg_ascend_down1}
\end{align}
for all $0 \leq j \leq m$.

The following lemma establishes the relations between $q$-degrees of the polynomial entries inside $\bF$.

\begin{prop}
	\label{lem:sorted_degrees}
	The following inequalities
	\begin{subequations}
		\label{eq:F_deg_ineq_across}
		\begin{align}
		\qdeg{F_{j'}(r',s)} &> \qdeg{F_{j}(r,s)}, \quad 0 \leq j < j' \leq m, \label{eq:f_deg_ascend_down2}\\ 
		\qdeg{F_{j'}(r,s')} &> \qdeg{F_{j}(r,s)}, \quad 0 \leq j < j' \leq m, \label{eq:f_deg_ascend_right2} 
		\end{align}
	\end{subequations}
	hold for all $0 \leq r, r' \leq n-1$ and $0 \leq s, s' \leq n-1$.
\end{prop}

\begin{IEEEproof}
	The inequality in~\eqref{eq:f_deg_ascend_down2} is established by
	\begin{align}
	&\qdeg{F_{j'}(r',s)} - \qdeg{F_{j}(r,s)} \nonumber\\
	&\quad= \qdeg{f_{s}^{[nj'+r']}} - \qdeg{f_s^{[nj+r]}}\nonumber \\
	&\quad= (nj'+r') + \qdeg{f_s} - (nj+r) - \qdeg{f_s}\nonumber \\								
	&\quad\geq n+r'-r \nonumber \\
	&\quad> 0,
	\end{align}
	where the first inequality follows from $j' \geq j+1$, and the second due to $r'-r > -n$.
	
	Using~\eqref{eq:f_deg_bounds}, we prove~\eqref{eq:f_deg_ascend_right2},
	\begin{align}
	&\qdeg{F_{j'}(r,s')} - \qdeg{F_{j}(r,s)} \nonumber\\
	&\qquad\quad= \qdeg{f_{s}^{[nj'+r]}} - \qdeg{f_s^{[nj+r]}}\nonumber \\
	&\qquad\quad\geq nj'+r - (n-1+nj+r) \nonumber \\
	&\qquad\quad> 0.
	\end{align}
	Here, the first inequality follows from~\eqref{eq:f_deg_bounds}, and the second due to $j' \geq j+1$.
\end{IEEEproof}

We note that~\eqref{eq:f_deg_ascend_down1},~\eqref{eq:f_deg_ascend_down2} and~\eqref{eq:f_deg_ascend_right2} parallel the inequalities~\eqref{eq:t_deg_ascend_down1},~\eqref{eq:t_deg_ascend_down2} and~\eqref{eq:t_deg_ascend_right2} for entries of $\bT$. 
It remains to find an equivalent relationship of \eqref{eq:t_deg_ascend_right1} for the entries of $\bF$.

To state the next property we let $\bC_i \in \field{F}_{q^M}^{n(m+1) \times n}$ and $\bR_i \in \field{F}_{q^M}^{n \times n(m+1)}$ be the $i$-th column and row blocks of $\bF$, i.e.,
\begin{align}
\bF = \begin{pmatrix}
\bC_0 , \bC_1 , \cdots , \bC_m
\end{pmatrix}
= \begin{pmatrix}
\bR_0 \\ \bR_1 \\ \vdots \\ \bR_m \\
\end{pmatrix}, \end{align}
\text{where} 
\begin{align} \bC_i \triangleq 
\begin{pmatrix}
\mathbf{0} \\ \vdots \\ \mathbf{0} \\ \bF_0 \\ \bF_1 \\ \vdots \\ \bF_i
\end{pmatrix} 
 \qquad \bR_i \triangleq 
\begin{pmatrix}
\mathbf{0} , \cdots , \mathbf{0} , \bF_0 , \bF_1 , \cdots , \bF_i
\end{pmatrix}.
\end{align}

The authors in \cite{ref:almeida_laa13} showed the following property. 
\begin{prop}
\label{prop:D_mat}
Let $\bD$ be an $l \times l$ sub-matrix of $\bF$ possessing a non-trivial determinant. 
The matrix $\bD$ can be written as
\begin{equation}
\label{eq:submatrix_structure}
\bD = 
\left(\begin{array}{c|ccc|c|c} 
\multicolumn{5}{c|}{\bO_h}	&	\multirow{5}{*}{$ \bD_h$} \\ \cline{1-5}
\multicolumn{4}{c|}{\bO_{h-1}}		&	\multirow{4}{*}{ $\bD_{h-1}$}	&	\\ \cline{1-4}
\multicolumn{4}{c|}{\vdots} 	&  \\ \cline{1-1}
\bO_{0}	&  	\multicolumn{3}{c|}{\multirow{2}{*}{$\dots$}}	& \\ \cline{1-1}
\bD_0	&	&	&	&	& 
\end{array}\right), 
\end{equation} 
where $0 \leq h \leq m$. 
Each $\bO_i$ is a zero matrix drawn from a single row block of $\bF$, whereas each $\bD_i$ is a matrix containing non-zero polynomials drawn from a single column block of $\bF$. 
Let $l_i \in \{1,\dots,n\}$ be the number of columns in each $\bD_i$ for $i \in \{0,\dots,h\}$. Then, $\sum_{i=0}^{h}l_i = l$. 
\end{prop}
\begin{rem}
In the special case of $h = m$, $\bD_i$ consists of columns drawn from the non-zero elements of $\bC_i$, whereas $\bO_i$ consists of rows drawn from the zero elements of $\bR_{m-i}$.
When $h < m$, it can be shown that $\bD$ preserves the structure in~\eqref{eq:submatrix_structure}.
If the $j$-th column block $\bC_j$ of $\bF$ was skipped when generating the submatrix $\bD$, one has to skip the $(m-j)$-th row block $\bR_{m-j}$ to avoid generating a trivial sub-matrix $\bD$. We refer the reader to~\cite{ref:almeida_laa13} for the complete proof.
\end{rem}

\begin{rem}
The degrees of the polynomials entries of $\bD$ satisfy~\eqref{eq:f_deg_ascend_down1},~\eqref{eq:f_deg_ascend_down2} and~\eqref{eq:f_deg_ascend_right2}.
\end{rem}

\begin{lem}
	\label{lem:M_existance}
	For $1 \leq l \leq n$, let $i_1,i_2,\dots,i_l$ be $l$ distinct elements of the set $\{0,\dots,n-1\}$. There exists a full-rank matrix $\bM \in \field{F}_{q}^{l \times l}$, such that the $q$-degrees of 
	$$(\hat{f}_{i_1},\hat{f}_{i_2},\dots,\hat{f}_{i_l}) = (f_{i_1},f_{i_2},\dots,f_{i_l}) \bM$$ 
	are monotonically increasing. 
	Moreover, the $q$-degrees of the polynomial entries of the product $(f^{[s]}_{i_1},f^{[s]}_{i_2},\dots,f^{[s]}_{i_l}) \bM$ are also sorted in the same order, for all $0 \leq s  \leq n(m+1)-1$.
\end{lem}
\begin{IEEEproof}
	As discussed in Remark~\ref{rem:independent_rows}, because $\alpha$ is a normal element of $\field{F}_{q^M}$ and $\bA$ is full rank, the polynomials $(f_{i_1},f_{i_2},\dots,f_{i_l})$ are linearly independent over $\field{F}_q$. 
	As a result, the isomorphic matrix $$\begin{pmatrix} \phi_n(f_{i_1}),\phi_n(f_{i_2}),\dots,\phi_n(f_{i_l}) \end{pmatrix} $$ is full column rank and can be transformed to reduced column echelon form $\begin{pmatrix} \phi(\hat{f}_{i_1}),\phi(\hat{f}_{i_2}),\dots,\phi(\hat{f}_{i_l}) \end{pmatrix}$ through elementary column operations, i.e., 
	\begin{align*}
	&\begin{pmatrix} \phi_n(\hat{f}_{i_1}),\phi_n(\hat{f}_{i_2}),\dots,\phi_n(\hat{f}_{i_l}) \end{pmatrix} \nonumber\\
	&\qquad\quad= \begin{pmatrix} \phi_n(f_{i_1}),\phi_n(f_{i_2}),\dots,\phi_n(f_{i_l}) \end{pmatrix} \bM,
	\end{align*}
	where $\bM \in \field{F}_{q}^{l\times l}$ is full rank matrix in the ground field. 
	This is equivalent to saying that the degrees of the polynomials $(\hat{f}_{i_1},\dots,\hat{f}_{i_l})$ are strictly increasing.
	
	The second part immediately follows, as $(f^{[s]}_{i_1},f^{[s]}_{i_2},\dots,f^{[s]}_{i_l}) \bM = (\hat{f}^{[s]}_{i_1},\hat{f}^{[s]}_{i_2},\dots,\hat{f}^{[s]}_{i_l})$ by using Freshman's rule.
\end{IEEEproof}

\subsection{Proof of Theorem~\ref{thm:esr}}

Having established the various properties in the previous section we are now in a position to complete the proof of Theorem~\ref{thm:esr}.

\begin{IEEEproof}[Proof of Theorem~\ref{thm:esr}]
Consider any nontrivial minor $\bD$ of $\bF$. We will show that $\det(\bD) \neq 0$.
First note from Property~\ref{prop:sorted_qdegs} and~\ref{lem:sorted_degrees} the elements of $\bD$ satisfy ~\eqref{eq:f_deg_ascend_down1},~\eqref{eq:f_deg_ascend_down2} and~\eqref{eq:f_deg_ascend_right2}, which are the counterparts of~\eqref{eq:t_deg_ascend_down1},~\eqref{eq:t_deg_ascend_down2} and~\eqref{eq:t_deg_ascend_right2} respectively.
We will show that $\bD$ can be transformed into a matrix $\hat{\bD}$ that satisfies the counterpart of~\eqref{eq:t_deg_ascend_right1}. 

 Using Property~\ref{prop:D_mat} we have that the structure of $\bD$ should satisfy~\eqref{eq:submatrix_structure}.
Consider the sub-matrices $\bD_0,\ldots, \bD_h$ associated with $\bD$.  
Each $\bD_j$ is contained entirely within a column block of $\bF$.
Thus it follows from~\eqref{eq:esr_blocks} that any row of $\bD_j$ can be expressed in the form $(f^{[s]}_{i_1},f^{[s]}_{i_2},\dots,f^{[s]}_{i_{l_j}})$ for some $0~\leq~s~\leq~n(m+1)-1$ and $0 \leq i_1< i_2 < \dots < i_{l_ j} \leq n-1$. 
In particular the indices $i_1, \ldots, i_{l_j}$ are common across all rows in $\bD_j$ and only the $q$-degree, denoted by $s$, varies across the rows.
Using Lemma~\ref{lem:M_existance}, we construct a full-rank matrix $\bM_j \in \field{F}_{q}^{l_j \times l_j}$, such that the degrees of the polynomials of $\hat{\bD}_j = \bD_j \cdot \bM_j$ will increase monotonically across the columns within each fixed row. Performing this for each $\bD_j$, we construct
	\begin{align}
	\hat{\bD} = \bD \cdot \bM,
	\end{align}
	where $\bM = \diag{(\bM_h,\dots,\bM_0)}$ is full-rank with elements in ${\mathbb F}_q$. It follows that the transformed matrix $\hat{\bD}$ satisfies the counterpart of~\eqref{eq:t_deg_ascend_right1}.

\begin{figure*}[!bh]
	\hrule
	\begin{align}
	\label{eq:long_f_equation}
	\bF &= \bT \bA_{[0,1]} =
	\begin{pmatrix}
	& & & & \alpha^{[0]} + \alpha^{[1]} & \alpha^{[1]} + \alpha^{[2]} + \alpha^{[3]} & \alpha^{[1]} & \alpha^{[0]} + \alpha^{[2]} \\
	& & & & \alpha^{[1]} + \alpha^{[2]} & \alpha^{[2]} + \alpha^{[3]} + \alpha^{[4]} & \alpha^{[2]} & \alpha^{[1]} + \alpha^{[3]} \\
	& & & & \alpha^{[2]} + \alpha^{[3]} & \alpha^{[3]} + \alpha^{[4]} + \alpha^{[5]} & \alpha^{[3]} & \alpha^{[2]} + \alpha^{[4]} \\
	& & & & \alpha^{[3]} + \alpha^{[4]} & \alpha^{[4]} + \alpha^{[5]} + \alpha^{[6]} & \alpha^{[4]} & \alpha^{[3]} + \alpha^{[5]} \\
	\alpha^{[1]} + \alpha^{[2]} & \alpha^{[0]} & \alpha^{[0]} + \alpha^{[2]} & \alpha^{[3]} & \alpha^{[4]} + \alpha^{[5]} & \alpha^{[5]} + \alpha^{[6]} + \alpha^{[7]} & \alpha^{[5]} & \alpha^{[4]} + \alpha^{[6]} \\
	\alpha^{[2]} + \alpha^{[3]} & \alpha^{[1]} & \alpha^{[1]} + \alpha^{[3]} & \alpha^{[4]} & \alpha^{[5]} + \alpha^{[6]} & \alpha^{[6]} + \alpha^{[7]} + \alpha^{[8]} & \alpha^{[6]} & \alpha^{[5]} + \alpha^{[7]} \\
	\alpha^{[3]} + \alpha^{[4]} & \alpha^{[2]} & \alpha^{[2]} + \alpha^{[4]} & \alpha^{[5]} & \alpha^{[6]} + \alpha^{[7]} & \alpha^{[7]} + \alpha^{[8]} + \alpha^{[9]} & \alpha^{[7]} & \alpha^{[6]} + \alpha^{[8]} \\
	\alpha^{[4]} + \alpha^{[5]} & \alpha^{[3]} & \alpha^{[3]} + \alpha^{[5]} & \alpha^{[6]} & \alpha^{[7]} + \alpha^{[8]} & \alpha^{[8]} + \alpha^{[9]} + \alpha^{[10]} & \alpha^{[8]} & \alpha^{[7]} + \alpha^{[9]} \\
	\end{pmatrix}
	\end{align}
	\end{figure*}

        We argue that the remaining conditions are also preserved under the transformation of $\bM$. Note that the elements on any given row of $\hat{\bD}_j$ are obtained through linear combinations (over ${\mathbb F}_q$)
        of the elements in the corresponding row of $\bD_j$. Since both~\eqref{eq:f_deg_bounds} and~\eqref{eq:Fjrs_def} are satisfied by all elements belonging to a fixed row of $\bD_j$, they must also be satisfied by $\hat{\bD}_j$.
        Then the proof of~\eqref{eq:f_deg_ascend_right2}, \eqref{eq:f_deg_ascend_down1} and~\eqref{eq:f_deg_ascend_down2} can be carried out for elements in $\hat{\bD}_j$.

	Consequently, $\hat{\bD}$ completely satisfies~\eqref{eq:T_deg_ineq_within} and~\eqref{eq:T_deg_ineq_across}, in contrast to $\bD$ which only satisfies~\eqref{eq:f_deg_ascend_down1} and~\eqref{eq:F_deg_ineq_across}. The proof showing that $\hat{\bD}$ is non-singular then follows verbatim to the proof showing all sub-matrices of $\bT$ are non-singular.
	Then, $\det{\hat{\bD}} = \det{\bD} \det{\bM}$ implies that $\bD$ is also non-singular and therefore, $\bF$ is super-regular.

	Recall for this proof, we fixed $\bA_0 = \bA_1 = \dots = \bA_m$. Without the assumption, the product $\bF$ would be comprised of a different set of $f_i(\alpha)$ for each column block. Because each $\bD_i$ is a sub-matrix from a single column block and column operations are performed for each $\bD_i$ independently, the polynomial degrees can always be transformed in order to satisfy~\eqref{eq:T_deg_ineq_within} and~\eqref{eq:T_deg_ineq_across}. Consequently, the proof is identical when considering a $\bA_{[0,m]}$ constructed from different blocks. 
\end{IEEEproof}

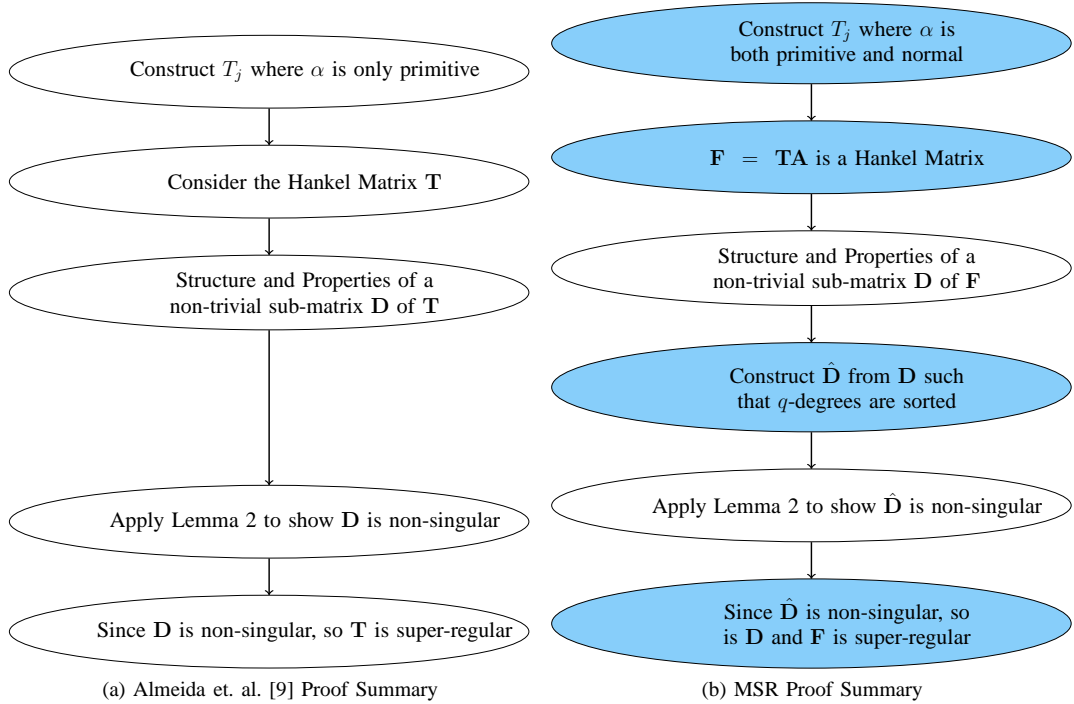
\begin{figure*}[ht!]
	\centering
	\subfloat[Almeida et. al.~\cite{ref:almeida_laa13} Proof Summary \label{subfig:Almeida_Proof_Summary}]
	{\resizebox{0.8\columnwidth}{!}
		{			
			\begin{tikzpicture}
			\node[myellipse={[20][4]} ,draw, fill = white] (e1) at (0em,0em) 			{Construct $T_j$ where $\alpha$ is only primitive};
			\node[myellipse={[20][4]} ,draw, fill = white] (e2) [below=2em of e1] 	{Consider the Hankel Matrix $\bT$};
			\draw[myarrow] (e1.south) -- (e2.north);
			\node[myellipse={[20][4]} ,draw, fill = white] (e3) [below=2em of e2] 	{Structure and Properties of a non-trivial sub-matrix $\bD$ of $\bT$};
			\draw[myarrow] (e2.south) -- (e3.north);
			\node[myellipse={[20][4]} ,draw, fill = white] (e5) [below=8.5em of e3] 	{Apply Lemma~\ref{lem:sr_prop} to show $\bD$ is non-singular};
			\draw[myarrow] (e3.south) -- (e5.north);
			\node[myellipse={[20][4]} ,draw, fill = white] (e6) [below=2em of e5] 	{Since $\bD$ is non-singular, so $\bT$ is super-regular};
			\draw[myarrow] (e5.south) -- (e6.north);
			\end{tikzpicture}
		}
	}
	\subfloat[MSR Proof Summary \label{subfig:MSR_Proof_Summary}]
	{\resizebox{0.8\columnwidth}{!}
		{			
			\begin{tikzpicture}
			\node[myellipse={[20][4]} ,draw] (e1) at (0em,0em) 			{Construct $T_j$ where $\alpha$ is both primitive and normal};
			\node[myellipse={[20][4]} ,draw] (e2) [below=2em of e1] 	{$\bF = \bT \bA$ is a Hankel Matrix};
			\draw[myarrow] (e1.south) -- (e2.north);
			\node[myellipse={[20][4]} ,draw, fill = white] (e3) [below=2em of e2] 	{Structure and Properties of a non-trivial sub-matrix $\bD$ of $\bF$};
			\draw[myarrow] (e2.south) -- (e3.north);
			\node[myellipse={[20][4]} ,draw] (e4) [below=2em of e3] 	{Construct $\hat{\bD}$ from $\bD$ such that $q$-degrees are sorted};
			\draw[myarrow] (e3.south) -- (e4.north);
			\node[myellipse={[20][4]} ,draw, fill = white] (e5) [below=2em of e4] 	{Apply Lemma~\ref{lem:sr_prop} to show $\hat{\bD}$ is non-singular};
			\draw[myarrow] (e4.south) -- (e5.north);
			\node[myellipse={[20][4]} ,draw] (e6) [below=2em of e5] 	{Since $\hat{\bD}$ is non-singular, so is $\bD$ and $\bF$ is super-regular};
			\draw[myarrow] (e5.south) -- (e6.north);
			\end{tikzpicture}
		}
	}
	\caption{A summary of the proof of Theorem \ref{thm:esr}, compared with the original proof given in \cite{ref:almeida_laa13}. The highlighted sections denote the key differences.}
	\label{fig:Proof_Summary}
\end{figure*}

We summarize the main steps in the proof of Theorem~\ref{thm:esr} as follows. A comparison of our proof and the original proof from \cite{ref:almeida_laa13} is provided in Fig.~\ref{fig:Proof_Summary}.
\begin{enumerate}
	\item We start by constructing the matrices $\bT_j$ in~\eqref{eq:sr_const} for $0 \leq j \leq m$ with $\alpha$ being a primitive normal over  $\field{F}_{q^M}$.
	\item We consider the product $\bF = \bT \bA_{[0,m]}$, where $\bT$ is the block Hankel matrix in~\eqref{eq:toeplitz_esr_rev} and $\bA_{[0,m]}$ is a block diagonal matrix in the ground field.   
	\item We next consider non-trivial sub-matrices $\bD$ of $\bF$ and study their structure and ordering of the $q$-degrees (Properties~\ref{prop:sorted_qdegs},~\ref{lem:sorted_degrees},~\ref{prop:D_mat}).
	\item In Lemma~\ref{lem:M_existance} we show that the matrix $\bD$ can be transformed into the matrix $\hat{\bD}$, where the $q$-degrees are sorted as required by Lemma~\ref{lem:sr_prop}.
	\item Applying Lemma~\ref{lem:sr_prop} it follows that $\hat{\bD}$ is non-singular. In turn this also implies that  $\bD$ is non-singular.
	\item This is equivalent to showing that $\bF$ is super-regular and the Theorem follows.
\end{enumerate}


\subsection{Example Illustrating the Proof of Theorem~\ref{thm:esr}}
\label{subsec:example_thm_esr}

In this subsection, we provide an example of the preservation of super-regularity, detailing the properties of $\bF$ and the key concepts in the proof of Theorem \ref{thm:esr}.

\begin{ex}
	\label{ex:thm_esr}
	Let $q = 2, n = 4, m = 1$ and $\alpha$ be a primitive normal element of $\field{F}_{2^{M}}$, where $M = q^{n(m+2)-1} = 2048$ satisfies Theorem~\ref{thm:esr}. Then,
	\begin{align*}
	{\bT} 	&= \begin{pmatrix} \mathbf{0} & \bT_0 \\ \bT_0 & \bT_1 \end{pmatrix} \\
	&= \begin{pmatrix}
	& & & & \alpha^{[0]} & \alpha^{[1]} & \alpha^{[2]} & \alpha^{[3]} \\
	& & & & \alpha^{[1]} & \alpha^{[2]} & \alpha^{[3]} & \alpha^{[4]} \\
	& & & & \alpha^{[2]} & \alpha^{[3]} & \alpha^{[4]} & \alpha^{[5]} \\
	& & & & \alpha^{[3]} & \alpha^{[4]} & \alpha^{[5]} & \alpha^{[6]} \\
	\alpha^{[0]} & \alpha^{[1]} & \alpha^{[2]} & \alpha^{[3]} & \alpha^{[4]} & \alpha^{[5]} & \alpha^{[6]} & \alpha^{[7]} \\
	\alpha^{[1]} & \alpha^{[2]} & \alpha^{[3]} & \alpha^{[4]} & \alpha^{[5]} & \alpha^{[6]} & \alpha^{[7]} & \alpha^{[8]} \\
	\alpha^{[2]} & \alpha^{[3]} & \alpha^{[4]} & \alpha^{[5]} & \alpha^{[6]} & \alpha^{[7]} & \alpha^{[8]} & \alpha^{[9]} \\
	\alpha^{[3]} & \alpha^{[4]} & \alpha^{[5]} & \alpha^{[6]} & \alpha^{[7]} & \alpha^{[8]} & \alpha^{[9]} & \alpha^{[10]} \\
	\end{pmatrix}
	\end{align*}
	is a super-regular matrix.

	Let $\bA_0$ and $\bA_1$ be the following two non-singular square matrices:
	\begin{align*}
	\bA_0 = \begin{pmatrix}
	0 & 1 & 1 & 0 \\
	1 & 0 & 0 & 0 \\
	1 & 0 & 1 & 0 \\
	0 & 0 & 0 & 1
	\end{pmatrix}, \quad
	\bA_1 = \begin{pmatrix}
	1 & 0 & 0 & 1 \\
	1 & 1 & 1 & 0 \\
	0 & 1 & 0 & 1 \\
	0 & 1 & 0 & 0
	\end{pmatrix}.
	\end{align*}
	
	We use $\bA_{[0,1]} = \diag{(\bA_0,\bA_1)}$ as the block-diagonal matrix and generate the product $\bF$  given in \eqref{eq:long_f_equation}.
	
	Because $\bA_0 \neq \bA_1$, the matrix $\bF$ is not a Hankel matrix. However, note that every block matrix posses a Moore structure. The elements in every row of a given block are Frobenius powers of the corresponding elements in the first row.
	Since $\alpha$ is a normal element in $\field{F}_{2^{2048}}$ and $\bA_0$ and $\bA_1$ are full rank, the isomorphism $\phi_n(\cdot)$ in Appendix~\ref{subsec:finite_fields} implies that the entries generating the left column block $ \alpha^{[1]} + \alpha^{[2]}, \alpha^{[0]}, \alpha^{[0]} + \alpha^{[2]}, \alpha^{[3]}$ are linearly independent polynomials. Similarly, the entries generating the right column block $\alpha^{[0]} + \alpha^{[1]}, \alpha^{[1]} + \alpha^{[2]} + \alpha^{[3]}, \alpha^{[1]}, \alpha^{[0]}+ \alpha^{[2]}$ are also respectively linearly independent amongst each other as stated in Remark~\ref{rem:independent_rows}.
	Note that the $q$-degree of these polynomials are upper and lower bounded between $3$ and $0$ which satisfies~\eqref{eq:f_deg_bounds}.
	
	Furthermore, the $q$-degrees are sorted according to Property~\ref{lem:sorted_degrees}.
	On a given column of $\bF$, the $q$-degrees are increasing by $1$ from top to bottom.
	Moreover, for any fixed row, any polynomial in the left block has a lower degree than any polynomial in the right block.
	
	Now consider the sub-matrix formed from rows $\br_i, i \in \{1,2,4,5 \}$ and columns $\bc_j, j \in\{3,4,5,6 \}$. This matrix
	\begin{align*}
	\bD =
	\begin{pmatrix}
	& \alpha^{[1]} + \alpha^{[2]}	& \alpha^{[2]} + \alpha^{[3]} + \alpha^{[4]}	& \alpha^{[2]} \\
	& \alpha^{[2]} + \alpha^{[3]}	& \alpha^{[3]} + \alpha^{[4]} + \alpha^{[5]}	& \alpha^{[3]} \\
	\alpha^{[3]} 	& \alpha^{[4]} + \alpha^{[5]}	& \alpha^{[5]} + \alpha^{[6]} + \alpha^{[7]}	& \alpha^{[5]} \\
	\alpha^{[4]} 	& \alpha^{[5]} + \alpha^{[6]}	& \alpha^{[6]} + \alpha^{[7]} + \alpha^{[8]}	& \alpha^{[6]} \\
	\end{pmatrix}
	\end{align*}
	does not have increasing $q$-degrees along rows, i.e., \eqref{eq:t_deg_ascend_right1} is not satisfied.
	The degree of the polynomials in the second and fourth columns are equal and lower than the degree of the polynomials in the third column. 
	The degrees can be sorted by subtracting the fourth column from the second and then swapping the third and the fourth columns.
	The transformed matrix
	\begin{align*}
	\hat{\bD} = \begin{pmatrix}
	& \alpha^{[1]} & \alpha^{[2]}	& \alpha^{[2]} + \alpha^{[3]} + \alpha^{[4]}	\\
	& \alpha^{[2]} & \alpha^{[3]}	& \alpha^{[3]} + \alpha^{[4]} + \alpha^{[5]}	\\
	\alpha^{[0]} 	& \alpha^{[4]} & \alpha^{[5]}	& \alpha^{[5]} + \alpha^{[6]} + \alpha^{[7]}	\\
	\alpha^{[1]} 	& \alpha^{[5]} & \alpha^{[6]}	& \alpha^{[6]} + \alpha^{[7]} + \alpha^{[8]}	\\
	\end{pmatrix}
	\end{align*}
	completely satisfies~\eqref{eq:T_deg_ineq_within} and~\eqref{eq:T_deg_ineq_across}. 
	The above column operations are equivalent to right multiplication with the matrix,
	\begin{align*}
	\bM = \begin{pmatrix}
	1	& 0	& 0	& 0 \\
	0	& 1 & 0	& 0 \\
	0	& 0 & 0 & 1 \\
	0	& 1	& 1 & 0 
	\end{pmatrix},
	\end{align*}
	which is full-rank. It follows that $\hat{\bD}$ is non-singular and therefore, $\bD$ as well.
\end{ex}

\section{Code Construction}
\label{sec:code_construct}

The rows of $\bT$ are permuted to form the block Toeplitz structure of an extended generator matrix
\begin{align}
\label{eq:toeplitz_esr_matrix}
\bar{\bT} = 
\begin{pmatrix}
\bT_0 & \bT_1 & \dots & \bT_m \\
	& \bT_0 & \dots & \bT_{m-1} \\
	&	& \ddots & \vdots \\
	&	&	& \bT_0
\end{pmatrix}.
\end{align}
Since every sub-matrix of $\bT$ has a counterpart in $\bar{\bT}$ identical up to row permutations, this block Toeplitz matrix is also super-regular.
$\bGEX_m$ is then constructed as a sub-matrix of $k(m+1)$ rows from $\bT$. This process parallels the construction of $m$-MDS generator matrices \cite{ref:smds_it06}. 

\begin{thm}
\label{thm:final_thm}
Let $\bar{\bT}$ be the super-regular matrix  in \eqref{eq:toeplitz_esr_matrix} generated using a primitive normal $\alpha \in \field{F}_{q^M}$, where $M = q^{n(m+2)-1}$. Let $0 \leq i_1 < \dots < i_k < n$ and construct a $k(m+1) \times n(m+1)$ sub-matrix $\bGEX_m$ of $\bar{\bT}$ from rows indexed $jn + i_1,\dots, jn+ i_k$ for $0 \leq j \leq m$. This matrix is the extended generator of an MSR convolutional code $\mathcal{C}[n,k,m]$.
\end{thm}

\begin{IEEEproof}
We show that $\bGEX_m$ satisfies Theorem \ref{thm:sum_rank_construct}. Assume without loss of generality that $i_1 = 0,\dots,i_k = k-1$. Each $\bT_i$ is divided into $\begin{pmatrix} \bG_i \\ \bT_i' \end{pmatrix}$, where $\bG_i \in \field{F}^{k\times n}_{q^M}$ are the blocks of the extended generator matrix. For $0 \leq t \leq m$, let $\bA_t \in \field{F}_q^{n \times n}$ be non-singular matrices. We similarly divide $\bA_t = \begin{pmatrix} \bA_t^* & \bA_t'   \end{pmatrix}$, where the two blocks $\bA_t^* \in \field{F}_q^{n\times \rho_t}$ and $\bA_t' \in \field{F}_q^{n\times (n-\rho_t)}$ represent the reduced channel matrix and some remaining matrix respectively.
Let $\bA_{[0,m]} = \diag{(\bA_0,\dots,\bA_m)}$. The product can be written as
\begin{align*} \bar{\bT} \bA_{[0,m]} =
\begin{pmatrix}
\bT_0 \bA_0 & \bT_1 \bA_1 & \dots & \bT_m \bA_m \\
		& \bT_0 \bA_1 & \dots	& \bT_{m-1} \bA_m \\
		&		& \ddots	& \vdots \\
		&		&		& \bT_0 \bA_m
\end{pmatrix},
\end{align*}
where 
\begin{align*}
\bT_i \bA_t = 
\begin{pmatrix}
\bG_i \bA_t^{*} & \bG_i \bA'_t \\
\bT'_i \bA_t^{*} & \bT'_i \bA'_t
\end{pmatrix}.
\end{align*}
The sub-matrix of $\bar{\bT} \bA_{[0,m]}$ containing only the rows and columns involving $\bG_i \bA_t^{*}$ is equal to the product $\bGEX_m \bA_{[0,m]}^*$. Because $\bar{\bT} \bA_{[0,m]}$ is super-regular, the determinant of $\bGEX_m \bA_{[0,m]}^*$ is either trivially zero or non-trivial and therefore non-zero. 

It can be shown that for all $\bA_{[0,m]}^*$ whose blocks satisfy \eqref{eq:thm_sum_rank_constraint}, the product
\begin{align*}
\bGEX_m \bA_{[0,m]}^* = 
\begin{pmatrix}
\bG_0 \bA_0^* & \bG_1 \bA_1^* & \dots & \bG_m \bA_m^* \\
		& \bG_0 \bA_1^* & \dots	& \bG_{m-1} \bA_m^* \\
		&		& \ddots	& \vdots \\
		&		&		& \bG_0 \bA_m^*
\end{pmatrix}
\end{align*}
has a non-trivial determinant. Note that the structure is reversed from that of $\bD$ in \eqref{eq:submatrix_structure}. In \cite{ref:almeida_laa13}, the authors showed that for $\bD$ to have a non-trivial determinant, the number of rows of each $\bD_j$ block cannot be less than the number of columns of each $\bO_j$ block. In this case each $\bD_j$ block of $\bGEX_m \bA_{[0,m]}^*$ contains $k(j+1)$ rows and each $\bO_j$ block contains $\sum_{t=0}^j \rho_t$ columns. Consequently if the condition in~\eqref{eq:thm_sum_rank_constraint} is satisfied, then $k(j+1) \geq \sum_{t=0}^j \rho_t$ for $j\leq m$ implies that $\bGEX_m \bA_{[0,m]}^*$ has a non-trivial determinant and is therefore non-singular.
Thus, $\bGEX_m$ satisfies Theorem \ref{thm:sum_rank_construct} and $\mathcal{C}[n,k,m]$ achieves $d_R(m) = (n-k)(m+1)+1$.
\end{IEEEproof}

\begin{rem}
Our construction thus far is feasible over any sliding window channel  $\isolatedchannel(S,W)$ with delay $T=W-1$ and $S < d_R(W-1)$. If instead $T \geq W$ holds, the same code $\mathcal{C}[n,k,T]$ is still feasible with the same threshold on $S$ i.e., $S < d_R(W-1)$.  However if $T<W-1$, then the code is feasible provided that $S < d_R(T)$.
\end{rem}


\subsection{Numerical Results}

The bound on the field size given in Theorem \ref{thm:esr} is only a sufficiency constraint required for the proof. For small code parameters, it is possible to numerically verify whether Theorem \ref{thm:sum_rank_construct} holds for a given extended generator matrix. An example is provided below of an MSR code over a small field. For the results in this section, we represent the primitive normal elements in the field as elements within the polynomial ring.

\begin{ex}
Let $\alpha = X+1$ be a primitive normal element in $\field{F}_{2^{11}} = \field{F}_2/\langle X^{11}+X^{2}+1 \rangle$. We construct the following extended generator matrix
\begin{align*}
	{\bGEX_1} 	= \begin{pmatrix}
	\alpha^{[0]} & \alpha^{[1]} & \alpha^{[2]} & \alpha^{[3]} & \alpha^{[4]} & \alpha^{[5]} & \alpha^{[6]} & \alpha^{[7]} \\
	\alpha^{[1]} & \alpha^{[2]} & \alpha^{[3]} & \alpha^{[4]} & \alpha^{[5]} & \alpha^{[6]} & \alpha^{[7]} & \alpha^{[8]} \\
	& & & & \alpha^{[0]} & \alpha^{[1]} & \alpha^{[2]} & \alpha^{[3]} \\
	& & & & \alpha^{[1]} & \alpha^{[2]} & \alpha^{[3]} & \alpha^{[4]}
	\end{pmatrix}.
	\end{align*}
This matrix satisfies Theorem \ref{thm:sum_rank_construct}, making it the generator for an MSR code $\mathcal{C}[4,2,1]$. Theorem \ref{thm:final_thm} guarantees the construction if $M=2^{11}$, i.e., $\alpha$ is a primitive normal element of  $\field{F}_{2^{2048}}$.
\end{ex}

\begin{ex}
Let $\alpha = X+1$ be a primitive normal element in $\field{F}_{2^{11}} = \field{F}_2/\langle X^{11}+X^{2}+1 \rangle$. The extended generator $\bGEX_2$ is given by
\begin{align*}
\begin{pmatrix}
\alpha		& \alpha^{[1]}		& \alpha^{[2]}		& \alpha^{[3]}		& \alpha^{[4]}		& \alpha^{[5]}		& \alpha^{[6]}		& \alpha^{[7]}		& \alpha^{[8]}	\\
\alpha^{[2]}	& \alpha^{[3]}		& \alpha^{[4]}		& \alpha^{[5]}		& \alpha^{[6]}		& \alpha^{[7]}		& \alpha^{[8]}		& \alpha^{[9]}		& \alpha^{[10]}	\\
			&				& 				& \alpha		& \alpha^{[1]}		& \alpha^{[2]}	& \alpha^{[3]}		& \alpha^{[4]}		& \alpha^{[5]}	\\	
			&				& 				& \alpha^{[2]}	& \alpha^{[3]}		& \alpha^{[4]}	& \alpha^{[5]}		& \alpha^{[6]}		& \alpha^{[7]}	\\	
			&				&				&				&				&						& \alpha		& \alpha^{[1]}		& \alpha^{[2]} \\
			&				&				&				&				&						& \alpha^{[2]}	& \alpha^{[3]}		& \alpha^{[4]}
\end{pmatrix}.
\end{align*}
This matrix satisfies Theorem \ref{thm:sum_rank_construct}, making it the generator for an MSR code $\mathcal{C}[3,2,2]$. Theorem \ref{thm:final_thm} guarantees the construction if $M=2^{11}$, i.e., $\alpha$ is a primitive normal element of  $\field{F}_{2^{2048}}$.
\end{ex}

The construction in Theorem \ref{thm:final_thm} can generate MSR codes over much smaller field sizes than than those given by the bound of $M \geq q^{n(m+2)-1}$. Table \ref{tab:achievable_small_fields} provides a list of code parameters and the field sizes on which they satisfy the matrix multiplication property of Theorem \ref{thm:sum_rank_construct}.

\begin{table*}
\centering
\begin{tabular}{c|c|c|c}
Parameters $[n,k,m]$ 	&	Field Bound		&	Achievable Field	& $\alpha$ 	\\ \hline \hline
$[4,2,1]$		&	$\field{F}_{2^{2048}}$	& $\field{F}_{2^{11}} = \field{F}_2/\langle X^{11}+X^{2}+1\rangle$	& $X+1$	\\
$[3,2,2]$		&	$\field{F}_{2^{2048}}$	& $\field{F}_{2^{11}} = \field{F}_2/\langle X^{11}+X^{2}+1\rangle$	& $X+1$	\\
$[3,1,2]$		&	$\field{F}_{2^{2048}}$	& $\field{F}_{2^{11}} = \field{F}_2/\langle X^{11}+X^{2}+1\rangle$	& $X+1$	\\
$[2,1,2]$		&	$\field{F}_{2^{128}}$		& $\field{F}_{2^{7}} = \field{F}_2/\langle X^{7}+X^{3}+1\rangle$	& $X^3+1$		\\
$[2,1,1]$		&	$\field{F}_{2^{64}}$		& $\field{F}_{2^{5}} = \field{F}_2/\langle X^{5}+X^{2}+1\rangle$	& $X+1$		
\end{tabular}
\caption{Achievable field sizes under which codes constructed using Theorem \ref{thm:final_thm}. The bound required by the theorem for each set of code parameters is provided in the middle column.}
\label{tab:achievable_small_fields}
\end{table*}

\section{Conclusion}

We introduce a new distance metric for convolutional codes called the column sum rank distance. We prove several properties  analogous to the column Hamming distance. A new family of codes --- MSR Codes --- that achieve the maximum distance up to the code memory is proposed. This is the rank metric counterpart to the $m$-MDS convolutional code. Our construction is based on matrices over extension fields that preserve super-regularity after multiplication with block diagonal matrices in the ground field.

The proof requires large field sizes but we numerically show that MSR codes do exist over smaller fields. Future work involves pursuing a more detailed study on field size requirements. Moreover, we have only considered a specific class of rank-deficient sliding window channels. In single-link streaming over burst erasure or mixed erasure channels, structured constructions using $m$-MDS codes as constituents have been revealed as more powerful alternatives \cite{ref:layered_it15}.  A similar study pertaining MSR codes remains an interesting direction of further study.


\appendices


\section{Mathematical Preliminaries}
\label{app:background}

\subsection{Finite Fields}
\label{subsec:finite_fields}

For $M \geq 0$ and a prime power $q$, let $\field{F}_q$ be the finite field with $q$ elements and $\field{F}_{q^M}$ be an extension field of $\field{F}_q$. A \textit{primitive} element $\alpha \in \field{F}_{q^M}$ is one whose consecutive powers can generate all non-zero elements of that field, i.e., $\field{F}_{q^M} = \{ 0, \alpha, \alpha^2,\dots, \alpha^{q^M-1} \}$. Let $\field{F}_q[X]$ be a polynomial ring of the ground field. The minimal polynomial of a primitive element $\alpha$ is the lowest degree monic polynomial $p_{\alpha}(X) \in \field{F}_q[X]$ for which $\alpha$ is a root. The minimal polynomial is irreducible and the degree of $p_{\alpha}(X)$ is equal to $M$. 
\begin{lem}
\label{lem:min_polynomial_divides}
If $f(\alpha) = 0$ for any $f(X) \in \field{F}_q[X]$, then $p_{\alpha}(X) \mid f(X)$.
\end{lem}
\begin{IEEEproof}
The proof can be found in \cite[Chapter~4]{ref:mcwilliams_text}.
\end{IEEEproof}

The extension field $\field{F}_{q^M}$ is isomorphic to the $M$-dimensional vector space $\field{F}_q^M$ over the ground field. Let $\alpha_0,\dots,\alpha_{M-1} \in \field{F}_{q^M}$ map to a basis for the vector space. A basis is defined as being \textit{normal} when for $0\leq i \leq m-1$, each $\alpha_i = \alpha^{q^i}$ for some $\alpha \in \field{F}_{q^M}$. The generating element $\alpha$ is referred to as a normal element. The notation $\alpha^{[i]} = \alpha^{q^i}$ is used to describe the $i$-th Frobenius power of $\alpha$. Every element $f \in \field{F}_{q^M}$ can be written as a linear combination in $\field{F}_q$ of the basis elements. Using the normal basis, $f$ resembles a linearized polynomial $f(X) = \sum_{i=0}^{M-1} f_i X^{[i]} \in \field{F}_q[X]$ evaluated at the normal element $X=\alpha$. The coefficients of this polynomial can be mapped
\begin{align}
\label{eq:isomorphism}
f(\alpha) = \sum_{i=0}^{M-1} f_i \alpha^{[i]}  & \mapsto \bff = (f_0,\dots,f_{M-1})^\transpose 
\end{align}
 to the entries of a unique vector $\bff \in \field{F}_q^{M \times 1}$. This mapping can be extended to vector spaces over the extension field and matrix spaces over the ground field. Using \eqref{eq:isomorphism}, we define $\phi_n :~ \field{F}_{q^M}^n \rightarrow \field{F}_q^{M\times n}$ as a bijection transforming a vector of linearized polynomial entries to a matrix whose columns are the coefficients of the polynomials. 

For every finite extension of a finite field, there exists at least one element that is both a normal and a primitive element \cite{ref:normal_basis_moc87}. Such an element is referred to as being \textit{primitive normal}. The properties of both normal and primitive elements are inherited in a primitive normal and will be useful in our code construction.

 A linearized polynomial is defined by the property that every monomial term must have a Frobenius power. A linearized polynomial possesses a $q$-degree, denoted as $\qdeg{f(X)}$, which gives the largest Frobenius power of the polynomial.

\subsection{Super-regular Matrices}
\label{subsec:bg_4}

For $b \in \field{N}$, let $\sigma$ be a permutation of the set $\left \{ 0,\dots,b-1 \right \}$. A permutation is comprised of a series of transpositions, which are defined as two entries of the set switching positions. The sign function of a permutation measures its parity, i.e., $\sgn{(\sigma)}$ is equal to $1$ when $\sigma$ is constructed from an even number of transpositions, and equal to $-1$ otherwise. Let $S_b$ denote the set of all possible permutations. The determinant of a $b\times b$  matrix $\bD$ can be calculated by summing over all permutations in $S_b$ in the Leibniz formula
\begin{align}
\label{eq:leibniz}
\det \bD \triangleq \sum_{\sigma \in S_b} \sgn{(\sigma)} \prod_{i=0}^{b-1} D_{i,\sigma(i)}.
\end{align}
Each product $\prod_{i=0}^{b-1} D_{i,\sigma(i)}$ is referred to as a term in the summation. When every term is equal to $0$, the matrix is said to have a trivial determinant. Using the Leibniz formula, a \textit{super-regular matrix} is a matrix for which every square sub-matrix with a non-trivial determinant is non-singular. 

\bibliographystyle{IEEEtran}

\begin{thebibliography}{10}
\providecommand{\url}[1]{#1}
\csname url@samestyle\endcsname
\providecommand{\newblock}{\relax}
\providecommand{\bibinfo}[2]{#2}
\providecommand{\BIBentrySTDinterwordspacing}{\spaceskip=0pt\relax}
\providecommand{\BIBentryALTinterwordstretchfactor}{4}
\providecommand{\BIBentryALTinterwordspacing}{\spaceskip=\fontdimen2\font plus
\BIBentryALTinterwordstretchfactor\fontdimen3\font minus
  \fontdimen4\font\relax}
\providecommand{\BIBforeignlanguage}[2]{{%
\expandafter\ifx\csname l@#1\endcsname\relax
\typeout{** WARNING: IEEEtran.bst: No hyphenation pattern has been}%
\typeout{** loaded for the language `#1'. Using the pattern for}%
\typeout{** the default language instead.}%
\else
\language=\csname l@#1\endcsname
\fi
#2}}
\providecommand{\BIBdecl}{\relax}
\BIBdecl

\bibitem{ref:mahmood_isit15}
R.~Mahmood, A.~Badr, and A.~Khisti, ``Convolutional codes in rank metric for
  network streaming,'' in \emph{IEEE Int. Symp. Inf. Theory (ISIT)}, 2015, to
  appear in 2015.

\bibitem{ref:martinian_it04}
E.~Martinian and C.-E.~W. Sundberg, ``Burst erasure correction codes with low
  decoding delay,'' \emph{IEEE Trans. Inf. Theory}, vol.~50, no.~10, pp.
  2494--2502, 2004.

\bibitem{ref:leong_isit13}
D.~Leong, A.~Qureshi, and T.~Ho, ``On coding for real-time streaming under
  packet erasures,'' in \emph{IEEE Int. Symp. Inf. Theory (ISIT)}, 2013, pp.
  1012--1016.

\bibitem{ref:layered_it15}
A.~Badr, P.~Patil, A.~Khisti, W.~Tan, and J.~Apostolopoulos, ``Layered
  constructions for low-delay streaming codes,'' \emph{IEEE Trans. Inf.
  Theory}, to appear in 2015.

\bibitem{ref:tomas_it12}
V.~Tomas, J.~Rosenthal, and R.~Smarandache, ``Decoding of convolutional codes
  over the erasure channel,'' \emph{IEEE Trans. Inf. Theory}, vol.~58, no.~1,
  pp. 90--108, 2012.

\bibitem{ref:ellis_pv12}
M.~Ellis, D.~P. Pezaros, and C.~Perkins, ``Performance analysis of {AL}-{FEC}
  for {RTP}-based streaming video traffic to residential users,'' in \emph{IEEE
  Packet Video Workshop}, 2012.

\bibitem{ref:smds_it06}
H.~Gluesing-Luerssen, J.~Rosenthal, and R.~Smarandache, ``Strongly-{MDS}
  convolutional codes,'' \emph{IEEE Trans. Inf. Theory}, vol.~52, no.~2, pp.
  584--598, 2006.

\bibitem{ref:hutch_laa08}
R.~Hutchinson, R.~Smarandache, and J.~Trumpf, ``Superregular matrices and the
  construction of convolutional codes having a maximum distance profile,''
  \emph{Lin. Algebra and App.}, vol. 428, pp. 2585--2596, 2008.

\bibitem{ref:almeida_laa13}
P.~Almeida, D.~Napp, and R.~Pinto, ``A new class of super regular matrices and
  {MDP} convolutional codes,'' \emph{Lin. Algebra and App.}, vol. 439, pp.
  2145--2157, 2013.

\bibitem{ref:ahlswede_it00}
R.~Ahlswede, N.~Cai, S.~Li, and R.~W. Yeung, ``Network information flow,''
  \emph{IEEE Trans. Inf. Theory}, vol.~46, no.~4, pp. 1204--1216, 2000.

\bibitem{ref:chou_allerton03}
P.~A. Chou, Y.~Wu, and K.~Jain, ``Practical network coding,'' in \emph{Allerton
  Conf. on Comm., Control, Comp.}, 2003.

\bibitem{ref:ho_it06}
T.~Ho, M.~M{\'e}dard, R.~K{\"o}tter, D.~R. Karger, M.~Effros, J.~Shi, and
  B.~Leong, ``A random linear network coding approach to multicast,''
  \emph{IEEE Trans. Inf. Theory}, vol.~52, no.~10, pp. 413--430, 2006.

\bibitem{ref:mrd_pit85}
E.~M. Gabidulin, ``Theory of codes with maximum rank distance,'' \emph{Probl.
  Inf. Transm.}, vol.~21, no.~1, pp. 1--12, 1985.

\bibitem{ref:roth_it91}
R.~M. Roth, ``Maximum-rank array codes and their application to crisscross
  error correction,'' \emph{IEEE Trans. Inf. Theory}, vol.~37, no.~2, pp.
  328--336, 1991.

\bibitem{ref:kotter_it08}
R.~K{\"o}tter and F.~R. Kschischang, ``Coding for errors and erasures in random
  network coding,'' \emph{IEEE Trans. Inf. Theory}, vol.~54, no.~8, pp.
  3579--3591, 2008.

\bibitem{ref:silva_it08}
D.~Silva, F.~R. Kschischang, and R.~K{\"o}tter, ``A rank-metric approach to
  error control in random network coding,'' \emph{IEEE Trans. Inf. Theory},
  vol.~54, no.~9, pp. 3951--3967, 2008.

\bibitem{ref:nobrega_winc10}
R.~W. N{\'o}brega and B.~F. Uch{\^o}a-Filho, ``Multishot codes for network
  coding using rank-metric codes,'' in \emph{IEEE Wireless Netw. Coding Conf.
  (WiNC)}, 2010, pp. 1--6.

\bibitem{ref:antonia_14}
A.~Wachter-Zeh, M.~Stinner, and V.~Sidorenko, ``Convolutional codes in rank
  metric with application to random network coding,'' \emph{IEEE Trans. Inf.
  Theory}, vol.~61, no.~6, pp. 3199--3213, 2015.

\bibitem{ref:khisti_isit10}
A.~Khisti, D.~Silva, and F.~R. Kschischang, ``Secure-broadcast codes over
  linear-deterministic channels,'' in \emph{IEEE Int. Symp. Inf. Theory
  (ISIT)}, 2010, pp. 555--559.

\bibitem{ref:jalali_ita11}
S.~Jalali, M.~Effros, and T.~Ho, ``On the impact of a single edge on the
  network coding capacity,'' in \emph{Inf. Theory and App. (ITA)}, 2011, pp.
  1--5.

\bibitem{ref:multicast_streaming_it15}
A.~Badr, D.~Lui, and A.~Khisti, ``Streaming codes for multicast over burst
  erasure channels,'' \emph{IEEE Trans. Inf. Theory}, vol.~61, no.~8, pp.
  4181--4208, 2015.

\bibitem{ref:zigangirov_fundamentals99}
R.~Johannesson and K.~S. Zigangirov, \emph{Fundamentals of Convolutional
  Coding}.\hskip 1em plus 0.5em minus 0.4em\relax IEEE Press, 1999.

\bibitem{ref:lin_ecc}
S.~Lin and D.~J. Costello, \emph{Error Control Coding}.\hskip 1em plus 0.5em
  minus 0.4em\relax Pearson Education Inc., 2004.

\bibitem{ref:gabidulin_acct88}
E.~M. Gabidulin, ``Convolutional codes over large alphabets,'' in \emph{Proc.
  Int. Workshop on Algebraic Combinatorial and Coding Theory (ACCT)}, Varna,
  Bulgaria, 1988, pp. 80--84.

\bibitem{ref:mahmood_phd}
R.~Mahmood, ``Rank metric convolutional codes with applications in network
  streaming,'' Master's thesis, University of Toronto, 2015.

\bibitem{ref:mcwilliams_text}
F.~J. MacWilliams and N.~J.~A. Sloane, \emph{Theory of Error-Correcting
  Codes}.\hskip 1em plus 0.5em minus 0.4em\relax North Holland Mathematical
  Library, 1977.

\bibitem{ref:normal_basis_moc87}
H.~W. Lenstra and R.~J. Schoof, ``Primitive normal bases for finite fields,''
  \emph{Math. Comp.}, vol.~48, no. 177, pp. 217--231, 1987.

\end{thebibliography}

\begin{IEEEbiographynophoto}
{Rafid Mahmood}
Rafid Mahmood received his B.A.Sc. and M.A.Sc. degrees in Electrical and Computer Engineering from the University of Toronto in Toronto, Ontario, Canada in 2013 and 2015 respectively. During this time, he foucsed his research on coding theory and information theory. He is pursuing his Ph.D. degree in Mechanical \& Industrial Engineering at the University of Toronto.
\end{IEEEbiographynophoto}
\vspace{-2em}
\begin{IEEEbiographynophoto}
{Ahmed Badr}
Ahmed Badr received the B.Sc., M.Sc. and Ph.D. degrees in Electrical \& Computer Engineering from Cairo University, Egypt in 2007, Nile University, Egypt in 2009 and University of Toronto, Canada in 2014. From September 2007 to August 2009, he was a Research Assistant in the Wireless Intelligent Networks Center (WINC), Nile University. In September 2009, he was a Research Assistant at the Signals Multimedia and Security Laboratory in University of Toronto. In 2014, he assumed his current position as a Postdoctoral Fellow in University of Toronto. His research interests include information theory, coding theory and real-time communication.
\end{IEEEbiographynophoto}
\vspace{-2em}
\begin{IEEEbiographynophoto}
{Ashish Khisti}
Ashish Khisti is an assistant professor and a Canada Research Chair in the Electrical and Computer Engineering (ECE) department at the University of Toronto, Toronto, Ontario Canada. He received his BASc degree in Engineering Sciences from University of Toronto and his S.M and Ph.D. Degrees from the Massachusetts Institute of Technology (MIT), Cambridge, MA, USA. His research interests span the areas of information theory, wireless physical layer security and streaming in multimedia communication systems. At the University of Toronto, he heads the signals, multimedia and security laboratory. He is a recipient of the Ontario Early Researcher Award from the Province of Ontario as well as a Hewlett-Packard Innovation Research Award.
\end{IEEEbiographynophoto}
\vfill

\end{document}